\documentclass[11pt,a4paper]{article}
\pdfoutput=1
\usepackage{jheppub}

\usepackage{ifpdf}
\usepackage{afterpage}
\usepackage{amssymb}
\usepackage{amsmath}
\usepackage{graphicx}
\usepackage{longtable}
\usepackage{verbatim}
\usepackage{amsfonts}
\usepackage{bbold}
\usepackage[table,usenames,dvipsnames]{xcolor}
\usepackage{titlesec}
\usepackage{enumerate}
\usepackage{gensymb}
\usepackage{cancel}

\usepackage{tikz}
\usetikzlibrary{trees}
\usetikzlibrary{decorations.pathmorphing}
\usetikzlibrary{decorations.markings}
\tikzset{
    photon/.style={decorate, decoration={snake}, draw=black},
    wino/.style={draw=redwine},
    electron/.style={draw=black, postaction={decorate},
        decoration={markings,mark=at position .55 with {\arrow[draw=black]{>}}}},
    scalar/.style={draw=black, dashed,postaction={decorate},
        decoration={markings,mark=at position .55 with {\arrow[draw=black]{>}}}},
    gluon/.style={decorate, draw=black,
        decoration={coil,amplitude=4pt, segment length=5pt}}
}

\newcommand{\UPQ}{U(1)_{\rm PQ}}

\newcommand{\bear}{\begin{array}}
\newcommand{\ear}{\end{array}}
\newcommand{\beq}{\begin{equation}}
\newcommand{\eeq}{\end{equation}}
\newcommand{\beqa}{\begin{eqnarray}}
\newcommand{\eeqa}{\end{eqnarray}}

\def\OMIT#1{{}}
\newcommand{\lsim}{\mathrel{\rlap{\lower4pt\hbox{\hskip1pt$\sim$}}
    \raise1pt\hbox{$<$}}}         
\newcommand{\gsim}{\mathrel{\rlap{\lower4pt\hbox{\hskip1pt$\sim$}}
    \raise1pt\hbox{$>$}}}         

\newcommand{\bl}{\left}
\newcommand{\br}{\right}

\newcommand{\mL}{\mathcal{L}}

\newcommand{\GeV}{\mathrm{GeV}}

\newcommand{\Eq}[1]{Eq.~(\ref{#1})}

\newcommand{\ignore}[1]{}

\vspace*{-30mm}

\title{\boldmath Challenges for an axion explanation of the muon $g-2$ measurement }
\author[a]{Manuel A. Buen-Abad,}
\author[a,b]{JiJi Fan,}
\author[c]{Matthew Reece,}
\author[d]{Chen Sun}
\affiliation[a]{Department of Physics, Brown University, Providence, RI, 02912, USA}
\affiliation[b]{Department of Physics and Brown Theoretical Physics Center, Brown University,
Providence, RI, 02912, U.S.A.}
\affiliation[c]{Department of Physics, Harvard University, Cambridge, MA, 02138, U.S.A}
\affiliation[d]{School of Physics and Astronomy, Tel-Aviv University, Tel-Aviv 69978, Israel}
\emailAdd{manuel\_buen-abad@brown.edu}
\emailAdd{jiji\_fan@brown.edu}
\emailAdd{mreece@g.harvard.edu}
\emailAdd{chensun@mail.tau.ac.il}

\vspace*{1cm}

\abstract{
The discrepancy between the muon $g-2$ measurement and the Standard Model prediction points to new physics around or below the weak scale. It is tantalizing to consider the loop effects of a heavy axion (in the general sense, also known as an axion-like particle) coupling to leptons and photons as an explanation for this discrepancy. We provide an updated analysis of the necessary couplings, including two-loop contributions, and find that the new physics operators point to an axion decay constant on the order of 10s of GeV. This poses major problems for such an explanation, as the axion couplings to leptons and photons must be generated at low scales. We outline some possibilities for how such couplings can arise, and find that these scenarios predict new charged matter at or below the weak scale and new scalars can mix with the Higgs boson, raising numerous phenomenological challenges. These scenarios also all predict additional contributions to the muon $g-2$ itself, calling the initial application of the axion effective theory into question. We conclude that there is little reason to favor an axion explanation of the muon $g-2$ measurement relative to other models postulating new weak-scale matter.
}

\begin{document}

\maketitle

\section{Introduction} 

The magnetic dipole moment of the muon is one of the most precisely measured quantities in particle physics. The discrepancy between theory and data is severe. The muon $g-2$ result, combining data from Brookhaven and Fermilab, is~\cite{Bennett:2006fi, Keshavarzi:2018mgv,NEWRESULT}
\beq
\Delta a_\mu = a_\mu^{\rm EXP} - a_\mu^{\rm SM} = (25.1 \pm 5.9) \times 10^{-10},
\eeq
where $a_\mu \equiv \frac{g-2}{2}$. Although the Standard Model calculation is fraught with difficulties, especially in pinning down hadronic contributions, various approaches have converged on consistent answers, as reviewed in Ref.~\cite{Aoyama:2020ynm}\footnote{However, see Ref.~\cite{Borsanyi:2020mff} for the BMW collaboration's most recent lattice computation of the hadronic vacuum polarization contribution to $a_\mu$, which disagrees with the data-driven value obtained by the Muon $g-2$ Theory Initiative \cite{Aoyama:2020ynm}.} (also, see the very recent lattice calculation of the hadronic light-by-light contribution, consistent with earlier estimates~\cite{Chao:2021tvp}). The muon $g-2$ anomaly is one of the most compelling discrepancies between theory and data that we have, so examination of possible new physics explanations are compelling. 

It is useful to put the $g-2$ anomaly in the context of other data. Any explanation of the new physics contribution to the anomalous muon magnetic dipole moment must contend with the {\em absence} of a signal in a variety of other experiments measuring lepton dipole operators.
The latest measurement of the electron $g-2$ is~\cite{Morel:2020dww}:
\beq
\Delta a_e = a_e^{\rm EXP} - a_e^{\rm SM}=(4.8 \pm 3.0) \times 10^{-13}. 
\eeq
Naively, in a model with minimal flavor violation, one would expect new physics to contribute\footnote{Notice that one factor of $m_e/m_\mu$ here comes from normalization of $a_{e,\mu}$, and one factor from the assumption that the dimension-six operator carries the appropriate lepton mass as the chiral symmetry violating parameter.} 
\begin{equation}
(\Delta a_e)_\mathrm{est.} \sim \left(\frac{m_e}{m_\mu}\right)^2 \Delta a_\mu \approx 5.9 \times 10^{-14},
\end{equation}
safely within  the range allowed by data. However, stringent constraints arise from measurements of CP- and flavor-violating operators. 
The current  bounds on the CP-violating electron EDM~\cite{Andreev:2018ayy} and muon EDM~\cite{Bennett:2008dy} are:
\begin{align}
|d_e| &< 1.1 \times 10^{-29}\,e\,\mathrm{cm}, \nonumber \\
|d_\mu| &< 1.9 \times 10^{-19}\,e\,\mathrm{cm}.
\end{align}
If the muon dipole operator has an imaginary part, it will produce a muon EDM; for an O(1) CP-violating phase, we would estimate
\begin{equation}
(d_\mu)_\mathrm{est} \sim \frac{e}{2m_\mu} \Delta a_\mu \approx 2.3 \times 10^{-22}\,e\,\mathrm{cm},
\end{equation}
safely below the experimental bound by a few orders of magnitude. However, if we naively rescale this to produce an electron EDM, it would be
\begin{equation}
(d_e)_\mathrm{est} \sim \frac{m_e}{m_\mu} (d_\mu)_\mathrm{est} \approx 1.1 \times 10^{-24}\,e\,\mathrm{cm},
\end{equation}
some {\em five orders of magnitude} above the experimental limit! From this, we learn that any putative explanation of the muon $g-2$ anomaly must provide a compelling reason why either CP violation is suppressed or the new physics couples much more strongly to muons than to electrons.
Charged lepton flavor violation is also highly constrained. The bound on the rare decay $\mu \to e \gamma$ is~\cite{TheMEG:2016wtm}:
\begin{equation}
\mathrm{Br}(\mu \to e\gamma) < 4.2 \times 10^{-13}.
\end{equation}
If we assume that the muon--electron dipole operator is of the same order as the anomalous muon magnetic dipole moment itself, we would have
\begin{equation}
\left[\mathrm{Br}(\mu \to e\gamma)\right]_\mathrm{est} \sim \frac{6\pi^2 e^2}{G_F^2 m_\mu^4} \left(\Delta a_\mu\right)^2 \approx 2.0 \times 10^{-3},
\end{equation}
which is {\em ten orders of magnitude} above the experimental bound. Hence, we need an extreme suppression of flavor-violating effects in any model that can explain the $g-2$ anomaly. Although the flavor puzzle at first glance appears to be numerically more severe than the CP puzzle, one should keep in mind that it involves a rate, which depends on the square of the coefficient of a dimension-six operator. In this sense, the two puzzles are comparable. Furthermore, we might expect that any extension of the Standard Model has approximate flavor symmetries, to explain the hierarchical quark and lepton masses as well as the pattern of small mixing angles in the quark sector. On the other hand, CP is badly violated in the CKM matrix, so the absence of CP phases demands an explanation.

Another puzzle that any new physics explanation of the muon $g-2$ discrepancy must confront is naturalness. One of the largest mysteries of the Standard Model is the electroweak hierarchy: why is the Higgs boson light, when quantum corrections tend to make scalar masses large? Most new physics explanations of the muon $g-2$ discrepancy invoke new scalar fields, which compound this problem, or new vector boson fields, which lead to similar problems if their masses arise from the Higgs mechanism. Although these new fields may be part of a consistent picture that addresses the hierarchy problem (for instance, when they are superpartners of Standard Model fields), it is interesting to ask if there are theories that address the muon $g-2$ anomaly that are relatively benign from the viewpoint of naturalness. New fermion fields can have technically natural small masses, so vectorlike leptons are one possible explanation (albeit one strongly constrained by data). Pseudo-Nambu-Goldstone bosons can also have naturally small masses. This leads us to the possibility of an {\em axion}, a light, periodic scalar boson arising from the breaking of an approximate $U(1)$ symmetry (known as a Peccei-Quinn or PQ symmetry)~\cite{Peccei:1977ur, Peccei:1977hh, Wilczek:1977pj, Weinberg:1977ma}, coupling to the muon as a possible explanation of the $g-2$ anomaly. Such an explanation has been studied in the past, often using an effective field theory (EFT) approach rather than a complete model~\cite{Chang:2000ii, Marciano:2016yhf, Bauer:2017ris, Bauer:2019gfk, Cornella:2019uxs, Chala:2020wvs, Bauer:2020jbp}. We use the word axion in the general sense, sometimes known as an axion-like particle, rather than assuming that the axion must solve the Strong CP problem. For a recent review of the phenomenology of such particles, see~\cite{Agrawal:2021dbo}.

In the Standard  Model, the dipole operators giving rise to the muon $g-2$ require a Higgs insertion to be gauge invariant, taking the form $H^\dagger L {\bar \sigma}^{\mu \nu} E^c B_{\mu \nu} + \mathrm{h.c.}$ or $H^\dagger \tau^i L {\bar \sigma}^{\mu \nu} E^c W^i_{\mu \nu} + \mathrm{h.c.}$, with coefficients that are complex in general and hence allow for CP violation. However, there are also operators like $L^\dagger {\bar \sigma}^\mu D^\nu L B_{\mu \nu}$, which do not violate chirality and become equivalent to the dipole operators only upon using equations of motion, which bring in factors of the Yukawa matrices. New physics that generates these non-chiral operators is thus an interesting possibility for contributing to muon $g-2$ without a corresponding electric dipole moment. A vectorlike lepton that mixes with the  muon is one possibility. Another is a derivatively-coupled axion, with an interaction like $(\partial_\mu a) (L^\dagger {\bar \sigma}^\mu L)$. One should still be wary of flavor off-diagonal couplings, which raise the prospect of dangerous contributions to $\mu \to e\gamma$. Nonetheless, we see that axions are interesting candidates to explain the $g-2$ anomaly for multiple reasons.

Having advocated for an axion as a promising explanation of the muon $g-2$ anomaly, we will now spend much of the paper arguing that this explanation, in reality, must overcome severe challenges. In Section~\ref{sec:explanation}, we provide a brief review of the effective theory of an axion, and then analyze the parameter space within this effective theory that is compatible with the observed value of the muon $g-2$ and unconstrained by other  experiments. Our analysis makes use of new two-loop calculations of Barr-Zee-type diagrams~\cite{Barr:1990vd} with derivative operator insertions, presented in Appendix~\ref{app:two-loop}. Importantly, we find that the characteristic mass scale suppressing the axion's couplings (its decay constant) is $\lesssim 25~\mathrm{GeV}$. This is a relatively low energy, at which we have exhaustive experimental probes of the Standard Model. Nonrenormalizable operators suppressed by this scale must be completed with particles whose masses are not much heavier. In Section~\ref{sec:models},  we examine some of the possible ultraviolet (UV) completions. In particular, derivative couplings of the axion to leptons suggest either that leptons carry PQ charge, which leads us to DFSZ-like models~\cite{Dine:1981rt,Zhitnitsky:1980tq}, or that they mix with new vectorlike leptons that carry PQ charge. UV completions predict one or more of: a relatively light radial mode of a PQ-charged, SM-neutral scalar (present in all UV completions), which in some models must mix with the Higgs boson; new charged fermions at around the weak scale, which might mix with ordinary leptons; and one or more additional, relatively light Higgs doublets. All of these new particles must evade stringent direct searches. A detailed analysis of constraints on UV completions is beyond the scope of this  paper, but by surveying various ways to generate the required EFT operators, we show that such models face severe challenges. Furthermore, in every case we find that the new particles and interactions required to complete the axion EFT lead to additional contributions to the muon $g-2$ itself, generally of the same order as  the contributions arising directly from the axion. Many of the ingredients that UV completions call for could also directly play a role in addressing the muon $g-2$ anomaly {\em without} needing to reconcile them with experimental constraints on axions. Thus, we conclude in Sec.~\ref{sec:con} that axions, on closer examination, have much dimmer prospects for explaining the anomaly than one might initially have hoped.

\section{Effective theory for the heavy axion $g-2$ explanation}
\label{sec:explanation}

One possible explanation for muon $g-2$ is to use heavy axion with mass above 30 MeV. Below this mass, axion couplings to photons are stringently constrained. In this section we will first review the general structure of the EFT for an axion-like particle, and then characterize the viable parameter space in which this EFT can lead to the observed $g-2$ anomaly. Readers wishing to bypass the theoretical preliminaries and go directly to our phenomenological effective Lagrangian and analysis of the parameter space can skip to Sec.~\ref{subsec:phenomenologicalLeff}. 

\subsection{Restrictions imposed by periodicity}
\label{subsec:periodicity}

We take our axion to be a periodic field, $a \cong a + 2 \pi f_a$, and refer to $f_a$ as the decay constant of the axion. The periodicity imposes an {\em exact} (gauged) discrete shift symmetry on the field, while also leading to an {\em approximate} continuous shift symmetry in the axion's interactions. In this subsection we will first offer general comments on the effective Lagrangian for such a periodic field; in the next subsection, we will consider additional constraints arising from the approximate continuous shift symmetry that is needed to keep the axion light. Finally, we will turn to phenomenological constraints on the resulting EFT. Our starting point will be somewhat different from that of earlier discussions of the axion EFT in~\cite{Georgi:1986df, Bauer:2020jbp, Chala:2020wvs, Bauer:2021wjo}, because we begin with only the assumption of periodicity. In particular, it is often stated that instanton effects break the axion's continuous shift symmetry to a discrete shift symmetry, but this is logically inverted; the axion begins its life as a periodic field, which constrains the form of its interactions with gauge fields. The literature on axion EFTs often invokes field redefinitions which are incompatible with the axion's periodicity. However, when we introduce simplifying assumptions about the form of our model, we will end up in a setting extremely close to that of the earlier references. (The earlier references also consider many details, like RG evolution in the axion EFT, that we do not.)

Working below the scale of electroweak symmetry breaking, we can package the charged lepton fields into Dirac fermion fields $\ell$ ($e, \mu,$ or $\tau$), with $\ell_L \equiv P_L \ell$ and $\ell_R \equiv P_R \ell$ denoting the left- and right-handed Weyl components. The effective Lagrangian containing the interactions relevant for our analysis takes the form:
\begin{align}
{\cal L}_{\rm eff} &\supset \frac{\partial_\mu a}{f_a} \left(\bar{\ell}_L k_L \gamma^\mu \ell_L + \bar{\ell}_R k_R \gamma^\mu \ell_R\right) +  \sum_{n \in \mathbb{Z}} \left(m^{jk}_n {\bar \ell}_{Rj} \mathrm{e}^{i n a/f_a} \ell_{Lk} + \mathrm{h.c.}\right) - V(a) \nonumber \\
& + c_{gg} \frac{\alpha_s}{8\pi} \frac{a}{f_a} G^a_{\mu\nu}\tilde{G}^{a\mu\nu} + c_{\gamma \gamma} \frac{\alpha}{4\pi} \frac{a}{f_a} F_{\mu\nu}\tilde{F}^{\mu\nu} + c_{\gamma\gamma;2} \frac{\alpha}{4\pi} \frac{\partial^2 a}{f_a^3} F_{\mu\nu}\tilde{F}^{\mu\nu} + \cdots.
\label{eq:effective}
\end{align}
Here $k_L$ and $k_R$ are hermitian matrices, $j$ and $k$  are lepton generation  numbers that are implicitly summed over, $F_{\mu\nu}$ is the electromagnetic field strength, $\tilde{F}^{\mu\nu}=\frac{1}{2} \epsilon^{\mu\nu \alpha\beta} F_{\alpha\beta}$, and $G^a_{\mu \nu}$ is the gluon field strength.\footnote{An even more general ansatz would allow the axion to couple to the lepton kinetic terms through a series of harmonics carrying $\exp(i n a/f_a)$ factors. We will briefly comment on this below.} Couplings to quarks could also exist, but will not be relevant for our analysis and take the same form as the lepton couplings, so we have suppressed them. The action must be invariant under the gauge transformation $a \mapsto a + 2\pi f_a$, which is highly constraining.\footnote{In theories of multiple periodic axions, naively integrating out heavier axions sometimes appears to give an EFT that need not respect shifts of only  the light axions. However, physical quantities computed in such EFTs, in known examples, still respect the requirements of periodicity, and we expect that a sufficiently careful choice of gauge and field redefinitions can always render the low-energy EFT manifestly periodic. See, e.g., Refs.~\cite{Choi:2019ahy,Fraser:2019ojt} for recent discussions along these lines. We will discuss related issues in Secs.~\ref{subsec:yuk-to-der} and~\ref{subsec:e-pq} below.} In particular, this is the reason for our choice to write the scalar $a {\bar \ell}_R \ell_L$ couplings in the form of a sum over integers $n$, manifestly respecting the periodicity. The $a F {\tilde F}$ term in the action is not gauge invariant, but must shift by a multiple of the periodicity of the QED $\theta$ term in order for $\exp(i S)$ to remain invariant. The conclusion depends on the global structure of the Standard Model gauge group, $(SU(3)_C \times SU(2)_L \times U(1)_Y)/\Gamma$. If $\Gamma = \mathbb{Z}_6$ or $\mathbb{Z}_3$, then $c_{\gamma\gamma}$ must be an integer, whereas if $\Gamma = \mathbb{Z}_2$ or $\mathbb{1}$, then $9 c_{\gamma \gamma}$ must be an integer~\cite{Tong:2017oea}. We will assume the former: $c_{\gamma\gamma} \in \mathbb{Z}$. Similarly, gauge invariance requires that $c_{gg} \in \mathbb{Z}$ and that $V(a)$ must be a periodic function with period $2\pi f_a$. Notice that derivative couplings like $c_{\gamma\gamma;2}$ are unconstrained by periodicity, and can give rise to effective contributions to $c_{\gamma\gamma}$ when the axion can be treated as massive, via the equation of motion $\partial^2 a = -\partial_a V(a) + \cdots \approx -m_a^2 a + \cdots$.

Notice that, because $c_{\gamma\gamma}$ and $c_{gg}$ are integers, it is {\em technically natural}  for an axion to not couple to photons or gluons (except via higher-derivative terms like $c_{\gamma\gamma;2}$). Such an assumption needs no more explanation than the fact that neutrinos are electrically neutral or that the electron does not carry color charge. Loops will induce these quantized couplings only when integrating out anomalous chiral fermions. Loops will generically induce the derivative couplings. For example, the loop-induced axion-photon couplings in Ref.~\cite{Bauer:2017ris} scale as $m_a^2$ in the light axion limit, because they arise from $c_{\gamma\gamma;2}$.

The above discussion has assumed that the gauge invariance $a \mapsto a + 2\pi f_a$ is not spontaneously broken. An alternative is that $V(a)$ (and analogously, the $a {\bar \ell}_R \ell_L$ couplings) may not be manifestly  periodic, but that periodicity is restored through the phenomenon of {\em monodromy}, with the potential having multiple branches~\cite{Silverstein:2008sg}. Our conclusions will only depend on the axion mass near the minimum of the potential, so the distinction will not be relevant for our purposes.

Above the weak scale, the axion can couple separately to $SU(2)_L$ and $U(1)_Y$, through $aW^i{\tilde W}^i$ and $aY {\tilde Y}$ couplings with quantized coefficients. However, one linear combination of these couplings can be removed by an anomalous $U(1)$ lepton-number transformation, leaving only the combination that becomes $c_{\gamma\gamma}$ in the low-energy EFT as physical~\cite{Tong:2017oea}. The removable linear combination can have physical effects only in the presence of explicit lepton-number violating interactions, like Majorana masses of neutrinos, where it will reappear in phases after the anomalous lepton-number transformation. The couplings of the axion to the $W$ and $Z$ will generally lead to only highly subdominant effects on the quantities that we study in the remainder of the paper, so we will ignore them.

\subsection{Restrictions imposed by approximate continuous shift symmetry}
\label{subsec:continuousshift}

The general Lagrangian Eq.~\eqref{eq:effective} is rather {\em too} general: on the one hand, it runs immediately into potential phenomenological problems from flavor and CP violation; on the other, it provides a much more general set of interactions than are found in typical UV completions. To some extent, these two considerations point toward the same simplifying assumptions.

One problem is that if multiple axion harmonics appear in the lepton mass terms, this can lead to CP violation and to substantial contributions to electric dipole moments. For example, if we have
\begin{equation}
\left(m_{e;0} + m_{e;1} \mathrm{e}^{i a/f_a}\right) {\bar e}_R e_L + \mathrm{h.c.}, \quad \arg(m_{e;0}) \neq \arg(m_{e;1}),
\end{equation}
then the relative phase of $m_{e;0}$ and $m_{e;1}$ will contribute to an electron EDM. We would like to eliminate these dangerous terms. Fortunately, in many UV completions, only a single harmonic appears. This is because, when the axion arises as a (pseudo)-Nambu-Goldstone boson of an approximate PQ symmetry~\cite{Peccei:1977ur, Peccei:1977hh, Wilczek:1977pj, Weinberg:1977ma}, only the harmonic $n = -\mathrm{PQ}(\ell_{Lk}) - \mathrm{PQ}({\bar \ell}_{Rj})$ is allowed by the underlying PQ symmetry (at least without further suppression due to PQ-violating effects). We can argue for this in more general language, based solely on the structure of the low-energy EFT: if the effective Lagrangian does not respect an approximate {\em continuous} symmetry that acts by shifting $a$, we would be free to write a potential with harmonics in $a$ multiplied by arbitrary mass scales of order the cutoff of the EFT, decoupling the axion entirely. A light axion requires an approximate symmetry. Nontrivial axion harmonics can appear in the fermion mass terms precisely when the fermions {\em also} transform under this approximate, continuous shift symmetry.

Thus, we assume that there is only a single harmonic in each of the terms in the sum and it is determined by the PQ charges of the lepton fields. This allows us to remove all axion couplings from the mass matrix, via field redefinitions of the form
\begin{equation}
\ell_{Lk} \mapsto \exp\left(-i \mathrm{PQ}(\ell_{Lk}) \frac{a}{f_a}\right) \ell_{Lk}, \quad {\bar \ell}_{Rj} \mapsto \exp\left(-i \mathrm{PQ}({\bar \ell}_{Rj}) \frac{a}{f_a}\right) {\bar \ell}_{Rk}.
\end{equation}
Because the PQ charges are assumed to be integers, these field redefinitions are compatible with the axion's periodicity. This chiral rotation of the lepton field's phases leads to a shift in the coupling $c_{\gamma\gamma}$ via the chiral anomaly. The derivative couplings also shift, acquiring new contributions arising from the field redefinition within the original lepton kinetic terms:
\begin{align}
(k_L)_{j,j'} &\mapsto (k_L)_{j,j'} + \mathrm{PQ}(\ell_{Lj}) \delta_{j,j'}, \nonumber \\
(k_R)_{k,k'} &\mapsto (k_R)_{k,k'} - \mathrm{PQ}({\bar \ell}_{Rk}) \delta_{k,k'}.
\end{align}
Here we have assumed, for simplicity, that the leptons' kinetic terms were originally diagonal. This assumption is not necessary. We could have started with a general lepton kinetic matrix including axion harmonics, restricted these harmonics to those corresponding to different PQ charges in off-diagonal terms, and carried out the same field redefinition. The result is an axion-independent kinetic matrix and a modified set of derivative couplings. We omit the details to avoid burying the reader in notation. The upshot of this discussion is that UV completions with a PQ symmetry can offer a good motivation to restrict our attention to the derivative couplings encoded by $k_L$ and $k_R$, and the couplings to gauge fields, while discarding the Yukawa-like couplings encoded by the $m^{jk}_n$ terms in Eq.~\eqref{eq:effective}. This brings us to an EFT of the form studied in Ref.~\cite{Georgi:1986df}, in which the light fermion fields do not transform under the PQ symmetry and the fermion-axion interactions occur only through derivative couplings that preserve a continuous shift symmetry.

\subsection{Phenomenological restrictions}
\label{subsec:phenomenologicalLeff}

At this point, we have gone as far as we can using only general properties of axions themselves. We expect the axion to couple to gauge fields (with quantized couplings, or through higher-derivative operators) and to have derivative couplings to fermions. The latter arise through hermitian matrices $k_L$ and $k_R$. These contain diagonal couplings which are real and flavor-conserving, but also off-diagonal couplings which are complex and flavor-violating. The latter couplings raise the possibility of severe constraints from precision tests of flavor and CP violation.

Indeed, there are strong constraints on flavor-violating derivative couplings of the axion to leptons, which have been studied recently in detail. The combination of an off-diagonal coupling $(k_R)_{e\mu}$ or $(k_L)_{e\mu}$ with a flavor conserving coupling like $(k_R)_{ii}, (k_L)_{ii}, c_{\gamma\gamma}$, or $c_{gg}$ can induce dangerous rates of charged lepton flavor violation processes like $\mu \to e\gamma$ or $\pi \to \mu e$~\cite{Bauer:2019gfk, Cornella:2019uxs, Dev:2017ftk}, as well as $\mu \to e a$ \cite{Calibbi:2020jvd}, which could be probed in the MEG-II experiment \cite{Baldini:2018nnn}. On the other hand, substantial off-diagonal couplings $(k_R)_{e\mu}$ or $(k_L)_{e\mu}$ {\em alone} can induce muonium-antimuonium oscillations~\cite{Endo:2020mev}.\footnote{Such dominantly off-diagonal couplings were entertained in Refs.~\cite{Bauer:2019gfk, Cornella:2019uxs} to explain both the muon $g-2$  anomaly and a discrepancy between the electron $g-2$ measurement~\cite{2008PhRvL.100l0801H, Hanneke:2010au} and a recent measurement of the fine structure constant~\cite{Parker:2018vye}. The latter anomaly is now in doubt given the latest fine structure constant measurement~\cite{Morel:2020dww}, and in any case, the purported explanation is excluded by muonium oscillations~\cite{Endo:2020mev}.}

In light of the strong experimental constraints on flavor-violating off-diagonal derivative couplings, we will henceforth assume that $k_L$ and $k_R$ are {\em diagonal} in the lepton mass basis. Equivalently, the flavor diagonal axion coupling is then\footnote{Note that using equation of motion or equivalently with chiral rotations of fermions, this operator could be rewritten as a combination~\cite{Bauer:2017ris, Bauer:2020jbp}:
  \beq
  \label{eq:da-to-a}
\frac{c_{ii}}{2} \frac{\partial_\mu a}{f_a} \bar{\ell}_i \gamma^\mu \gamma_5 \ell_i = - c_{ii} \frac{m_i}{f_a} a\bar{\ell}_i  i \gamma_5 \ell_i + c_{ii}\frac{\alpha}{4\pi} \frac{a}{f_a} F_{\mu\nu}\tilde{F}^{\mu\nu} + \cdots,
\eeq
where the dots represent similar terms involving $Z$ boson and terms higher order in $a/f_a$.} 
\beq
\frac{c_{ii}}{2}  \frac{\partial_\mu a}{f_a} (\bar{\ell}_i \gamma^\mu \gamma^5 \ell_i), \quad c_{ii} = (k_R)_{ii} - (k_L)_{ii}. 
\label{eq:fermioncoupling}
\eeq
Our focus is not the QCD axion, and the coupling to gluons will not be important for our phenomenological analysis. Furthermore, as mentioned above, it is technically natural to set it to zero. Hence, we will neglect $c_{gg}$ as well. (We will also mostly neglect couplings to quarks, which may exist but do not affect our observables of interest.) We are left with only the diagonal derivative couplings~\eqref{eq:fermioncoupling} and the couplings to photons. Summarizing, then, the axion EFT that we will work with in the remainder of the paper has the form
\begin{align}
{\cal L}_{\rm eff} &\supset \frac{c_{ii}}{2}  \frac{\partial_\mu a}{f_a} (\bar{\ell}_i \gamma^\mu \gamma^5 \ell_i) - V(a) + c_{\gamma \gamma} \frac{\alpha}{4\pi} \frac{a}{f_a} F_{\mu\nu}\tilde{F}^{\mu\nu} + c_{\gamma\gamma;2} \frac{\alpha}{4\pi} \frac{\partial^2 a}{f_a^3} F_{\mu\nu}\tilde{F}^{\mu\nu} + \cdots.
\label{eq:effective2}
\end{align}
This EFT can be valid up to energy scales $\lesssim 4\pi f_a$. We emphasize that the truncation to only flavor-conserving couplings is not justified on general EFT grounds in the infrared. An ultraviolet completion must have some structure, such as flavor symmetries, to explain the suppression of the flavor-violating off-diagonal couplings. Because our main goal is to argue that axion explanations of the muon $g-2$ anomaly face a variety of challenges in their UV completion, this only strengthens our main conclusion.

\subsection{Parameter space for explaining muon $g-2$}
\label{subsec:parameterspace}

\begin{figure}[!h]
\centering
\includegraphics[width=1.0\textwidth]{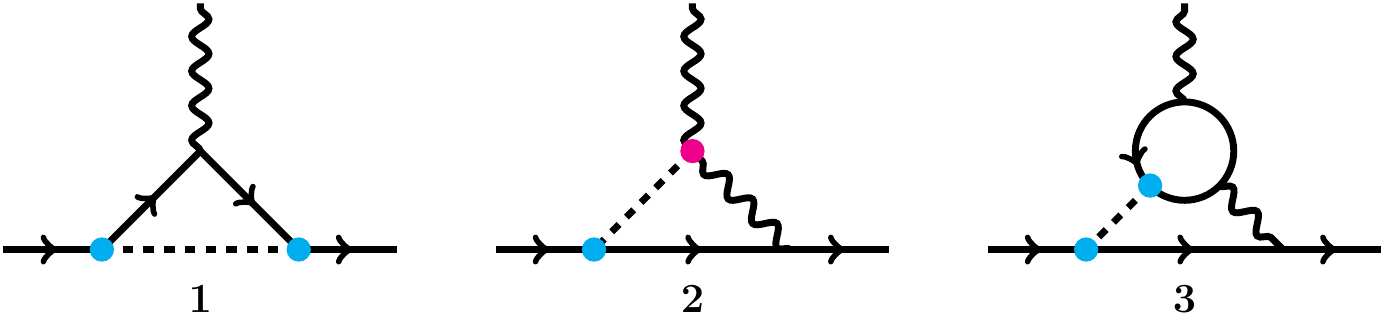}
\caption{Feynman diagrams for contributions to lepton $g-2$. The cyan dots represent insertions of a derivative coupling of the form~\eqref{eq:fermioncoupling}. The magenta dot represents an insertion of an $a F {\tilde F}$ coupling. Unmarked vertices are ordinary gauge interactions.}
\label{fig:diagram}
\end{figure}

In this effective Lagrangian, there are three leading diagrams contributing to the lepton $g-2$, as depicted in Fig.~\ref{fig:diagram}.

Schematically, the contributions of these three diagrams to $a_\mu$ are 
\beq
\Delta a_\mu^{(1)} \propto - \frac{c_{\mu\mu}^2}{16\pi^2}, \quad  \Delta a_\mu^{(2)} \propto - \frac{c_{\mu\mu}c_{\gamma\gamma} \alpha}{16\pi^3}, \quad \Delta a_\mu^{(3)} \propto - \frac{c_{\mu\mu}c_{ii} \alpha}{16\pi^3},
\eeq
where for diagram (3), $c_{ii}$ is the axion coupling to a fermion $\ell_i$ with a mass $m_i$ running in the loop. 
Note that the first diagram always has the wrong-sign contribution to muon $g-2$. The remaining two contributions actually contribute at the same order (due to the fact that $c_{\gamma\gamma}$ always comes with a one-loop factor). They could have the correct sign if $c_{\mu\mu}$ and $c_{\gamma\gamma}$ (or $c_{ii}$) have opposite signs. Thus to explain the muon $g-2$ anomaly, we need to have at least one of the latter two diagrams to balance against the first one. This possibility has been discussed in Refs.~\cite{Chang:2000ii, Marciano:2016yhf, Bauer:2017ris}.

One technical subtlety, which is not fully addressed in the literature, is a direct calculation of diagram (3) with fermions of all possible masses running in the inner loop. \footnote{A calculation of the two-loop contribution to electron $g-2$ was carried out in a different operator basis and then transformed to our basis in Ref.~\cite{Buttazzo:2020vfs}, which result agrees with ours.} In Refs.~\cite{Bauer:2017ris, Bauer:2020jbp}, the vertex function from the fermion loop with the axion and the two photons all on mass shell has been computed first and then inserted into diagram (2) to get an approximated answer for the third diagram. Yet the more proper treatment is as follows: {\it a)} compute the fermion loop contribution to the vertex function with only one photon on-shell and do not impose the on-shell conditions for the axion and the other photon; {\it b)} insert the vertex function into diagram (2) to get the final answer.\footnote{Ref.~\cite{Chang:2000ii} did a full two-loop calculation with the operator $a \bar{\ell} i \gamma_5 \ell$, following this strategy but assuming the fermions in the loop being heavy. We also want to consider the case with a fermion lighter than the muon, i.e., an electron, running in the loop.} With the shift-invariant axion-fermion coupling in Eq.~\eqref{eq:fermioncoupling}, we follow the recipe above and perform a two-loop calculation for diagram (3). The full results are included in App.~\ref{app:two-loop}. In two interesting limits, we have 
\beqa
a_\mu^{(3)} &\approx & - \frac{c_{\mu\mu}c_{ii}\alpha }{8\pi^3} \frac{m_\mu^2}{f_a^2}\, \ln \left(\frac{\Lambda^2}{m_i^2}\right), \quad \quad \quad  m_\mu  \ll m_a \ll m_i  \ll  \Lambda, \nonumber \\
&\approx & -  \frac{c_{\mu\mu}c_{ii}\alpha }{8\pi^3} \frac{m_\mu^2}{f_a^2}\, \left(\ln \frac{\Lambda^2}{m_a^2}+2\right), \quad  m_i \ll m_\mu \ll m_a \ll \Lambda,
\eeqa
where $\Lambda$ is the UV cutoff scale. 
Note that in the limit $m_\mu \ll m_i \ll \Lambda$, the contribution is from a heavy fermion loop and one would expect that it should decouple. The result above is from the renormalization of $a_\mu$ due to the axion-fermion couplings at the two-loop order between $m_i$ and $\Lambda$. The formula for diagrams (1) and (2) have been computed in Ref.~\cite{Marciano:2016yhf, Bauer:2017ris} and we will include them in App.~\ref{app:one-loop}. 

\begin{figure}[h]
\centering
\includegraphics[width=0.45\textwidth]{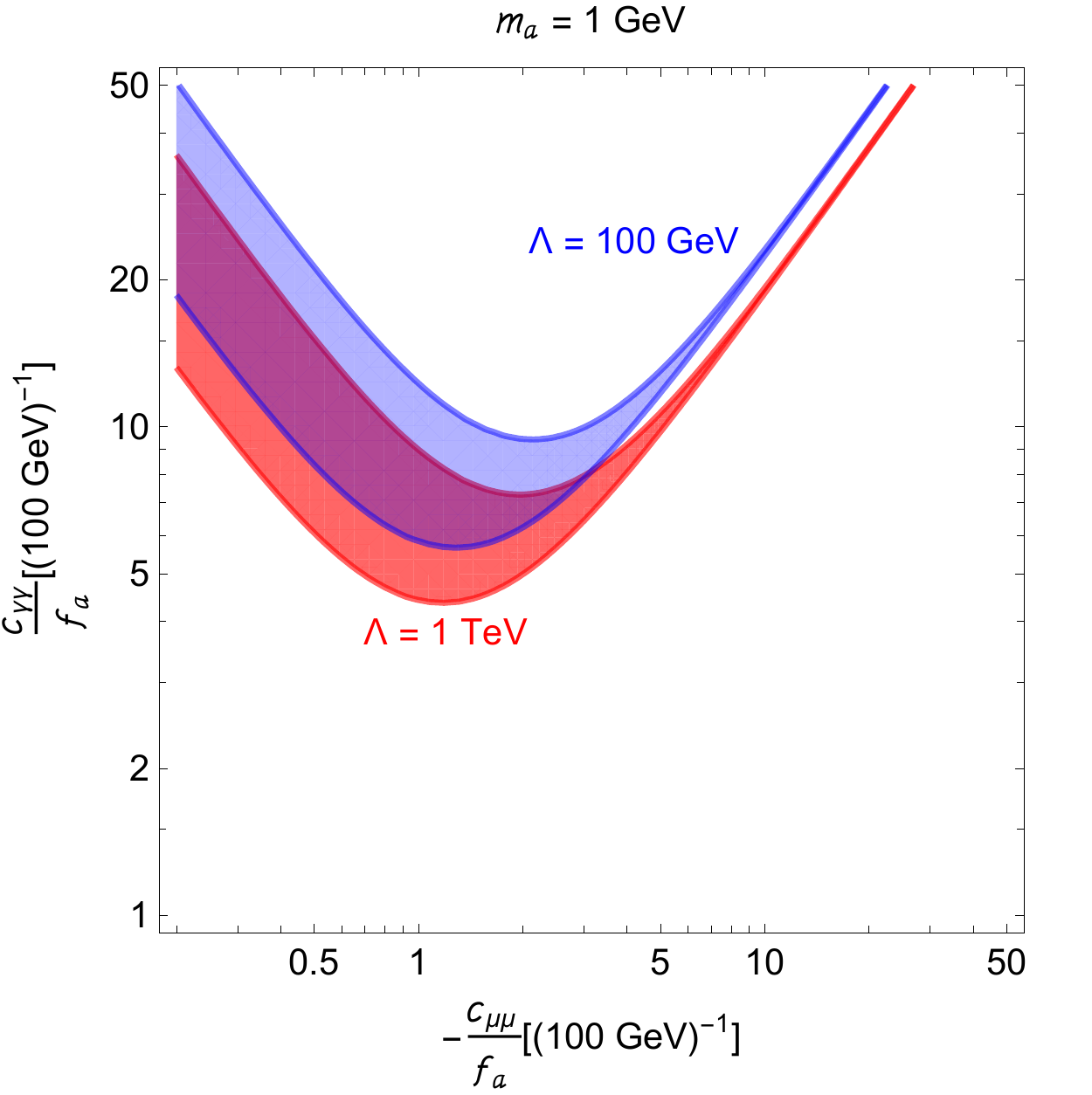} \quad  \includegraphics[width=0.45\textwidth]{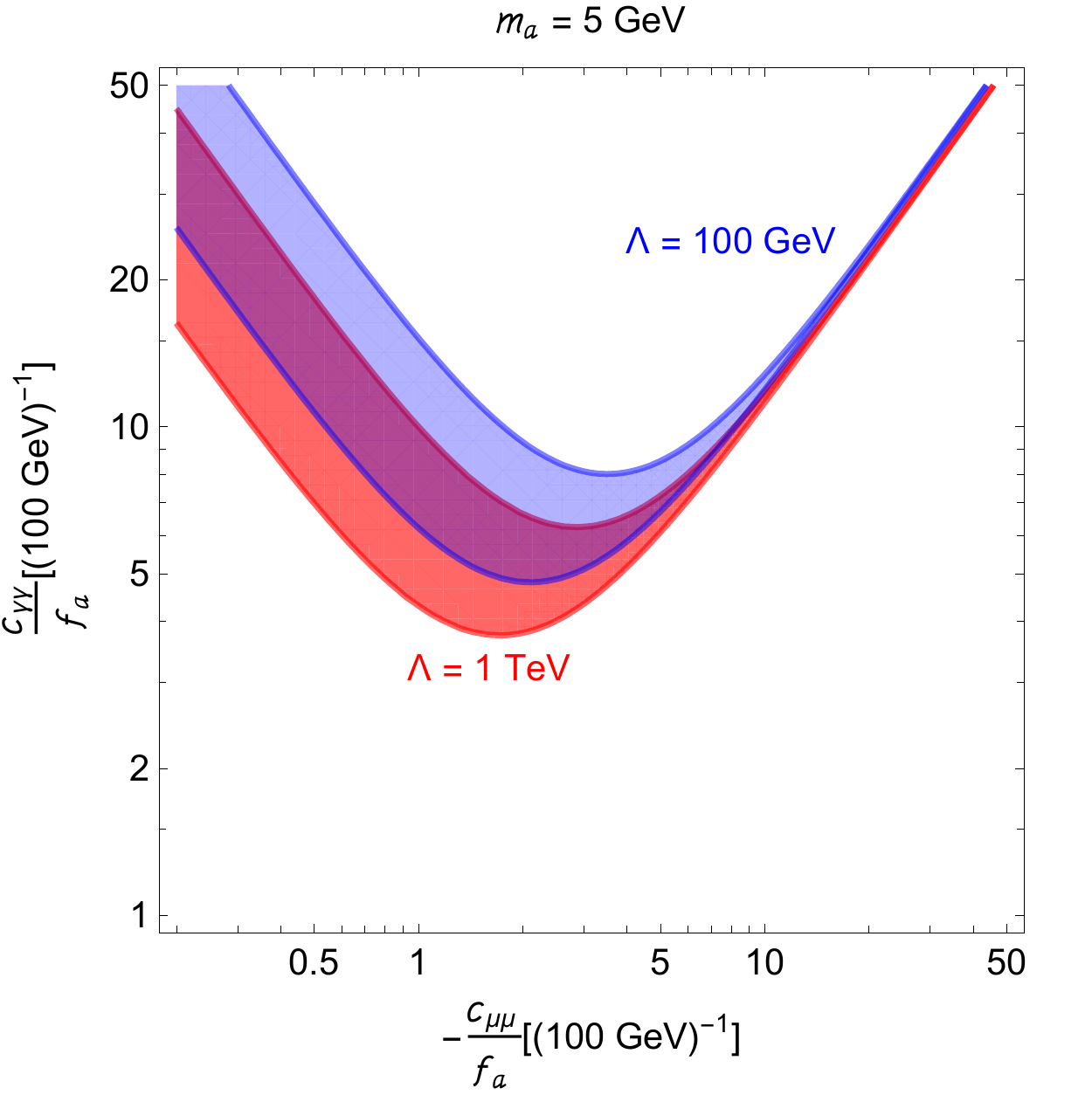}
\caption{Allowed regions (red) in the ($c_{\mu\mu}/f_a$, $c_{\gamma\gamma}/f_a$) plane to explain muon $g-2$ at $2\sigma$ level, for $m_a = 1$ GeV (left) and $m_a = 5$ GeV (right). We set all the other axion couplings to be zero. The red band corresponds to cutoff scale $\Lambda = 1$ TeV while the blue band corresponds to $\Lambda = 100$ GeV.}
\label{fig:photoncoupling}
\end{figure}

\begin{figure}[h]
\centering
\includegraphics[width=0.6\textwidth]{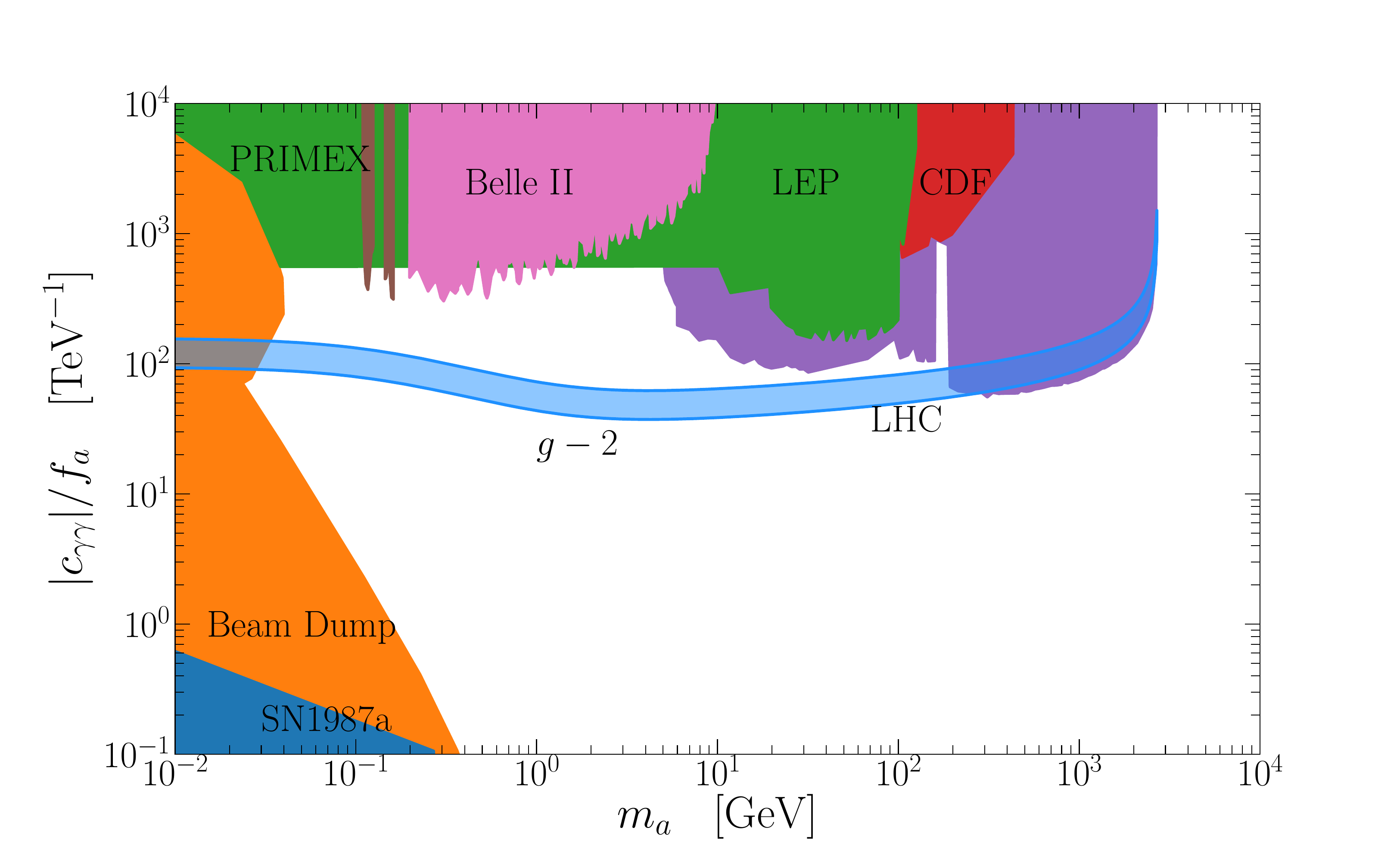}
\includegraphics[width=0.6\textwidth]{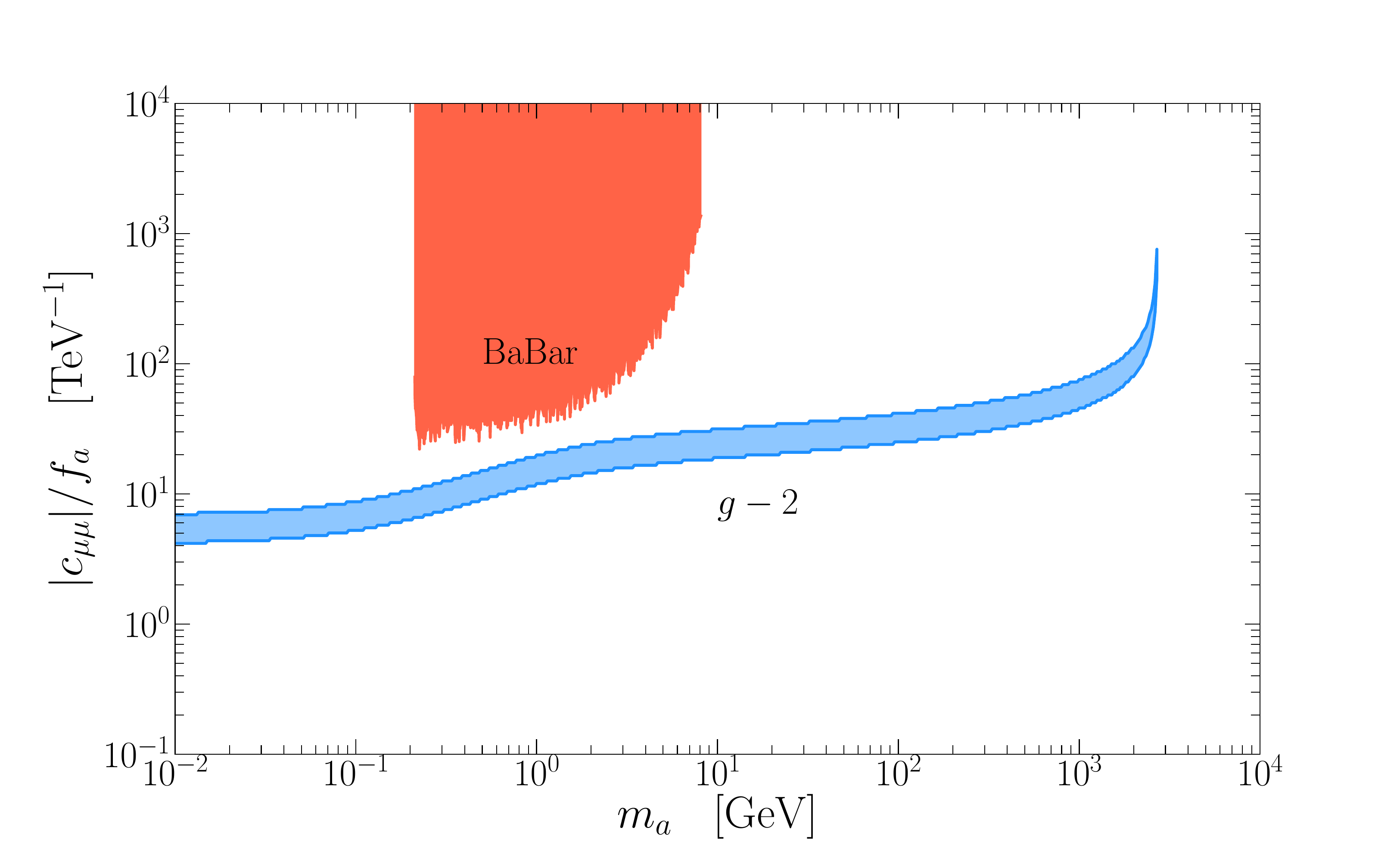}
\caption{Top: minimal $c_{\gamma\gamma}/f_a$ needed to explain the muon $g-2$ anomaly as a function of $m_a$ (upper blue band), varying $c_{\mu\mu}$. The other colored regions are the current experimental constraints, taken from Fig.~29 of Ref.~\cite{Agrawal:2021dbo}, as well as the recent bounds from PRIMEX \cite{Aloni:2019ruo} and Belle II \cite{BelleII:2020fag}. Note that for the LHC bound, which comes from $a \to \gamma\gamma$ channel, we need to take into account the contribution of the axion-muon coupling to both the production of axions and their branching fraction into photons. Bottom: corresponding $c_{\mu\mu}/f_a$ when $c_{\gamma\gamma}/f_a$ is minimized.
The BaBar constraint is a recast of their $e^+ e^- \to \mu^+ \mu^- Z'$ bound in~\cite{TheBABAR:2016rlg} using FeynRules~\cite{Alloul:2013bka} and MadGraph~\cite{Alwall:2014hca}, assuming the given $c_{\mu\mu}$ and (for the purposes of obtaining a branching fraction) the minimal $c_{\gamma\gamma}$ needed to explain the muon $g-2$ anomaly. In practice, the muonic branching fraction is nearly 100\% in the entire excluded region. In both panels, we take $\Lambda = 1$ TeV. }
\label{fig:massphotoncoupling}
\end{figure}

\begin{figure}[h]
\centering
\includegraphics[width=0.45\textwidth]{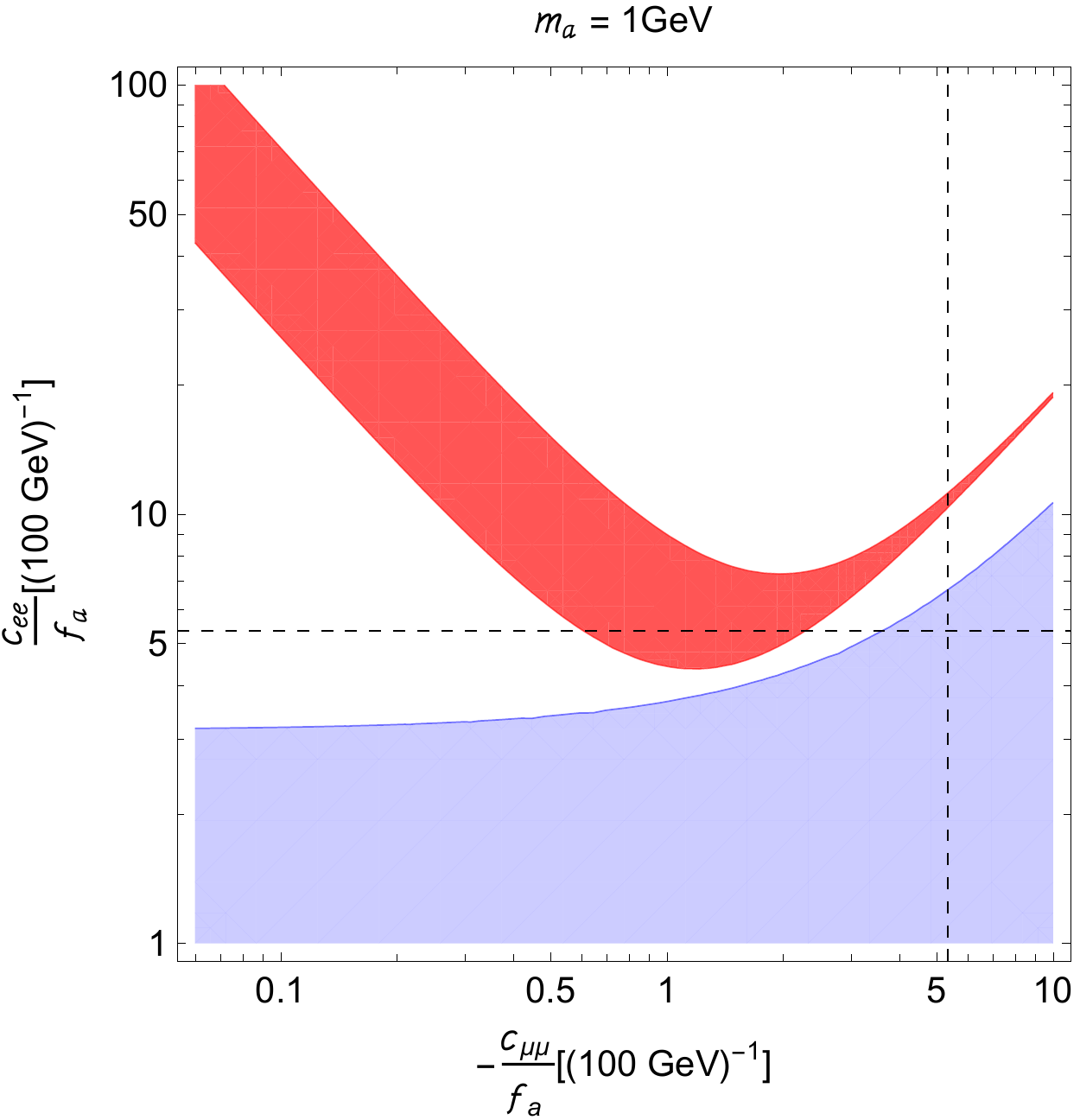} \quad \includegraphics[width=0.45\textwidth]{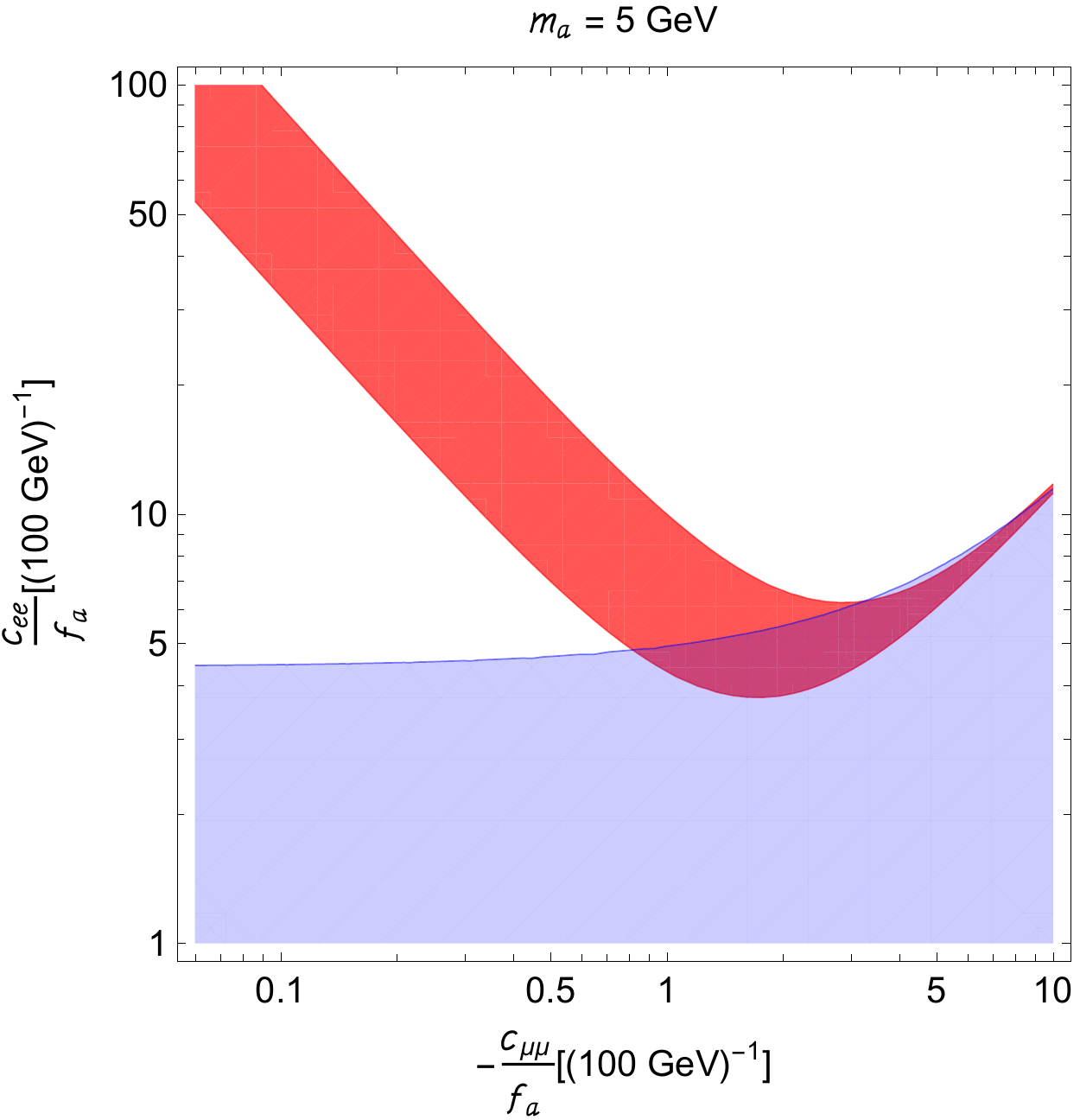}
\caption{Regions compatible with muon $g-2$ data (red) and electron $g-2$ (blue) at 2$\sigma$ level, in the ($c_{\mu\mu}$, $c_{ee}$) plane. Left: $m_a = 1$ GeV; right: $m_a = 5$ GeV. In the left plot, the black dashed lines are lower bounds on the axion-lepton couplings, derived from $e^+e^- \to 4 \mu$ search at BaBar~\cite{TheBABAR:2016rlg}. We set all the other axion couplings to be zero. In both panels, we take $\Lambda = 1$ TeV.}
\label{fig:muonelectron}
\end{figure}

To be more quantitative, we consider two scenarios below. 
\begin{itemize}
\item If we only include non-zero $c_{\mu\mu}$ and $c_{\gamma\gamma}$ (all three diagrams in Fig.~\ref{fig:diagram} contribute), then the parameter space to explain muon $g-2$ for $m_a = 1$ and 5 GeV is in Fig.~\ref{fig:photoncoupling}. Allowing $c_{\mu\mu}$ to vary, we show the minimal $c_{\gamma\gamma}/f_a$ needed to explain the muon $g-2$ anomaly as a function of $m_a$ in Fig.~\ref{fig:massphotoncoupling}. The allowed parameter space, consistent with current experimental bounds, is then
\beqa
c_{\gamma\gamma}/c_{\mu\mu} &<& 0,   \quad m_a \subset (40 \, {\rm MeV} - 200 \, {\rm GeV})   \nonumber \\
 \left|\frac{f_a}{c_{\gamma\gamma}} \right| & \lesssim & (10 - 25) \, {\rm GeV}, \quad \left|\frac{f_a}{c_{\mu\mu}} \right|  \lesssim 100 \, {\rm GeV} \ .
    \label{eq:params1}
\eeqa
\item If we only include non-zero $c_{\mu\mu}$ and $c_{ee}$ (contributions from diagram (1) and (3) in Fig.~\ref{fig:diagram}), the parameter space that is consistent with observed muon and electron $g-2$ values is shown in Fig.~\ref{fig:muonelectron}.\footnote{See Ref.~\cite{Darme:2020sjf} for previous work in the context of axion-like particles as portals to dark sectors, with older measurements of the electron $g-2$.} From it, we could see that
\beqa
&&m_a \gtrsim  2 \, {\rm GeV}, \quad c_{ee}/c_{\mu\mu}<0, \nonumber \\
&&\left|\frac{f_a}{c_{\mu\mu}}\right| \lesssim  100 \,{\rm GeV}, \quad  \left|\frac{f_a}{c_{ee}}\right| \lesssim  25\,{\rm GeV} \quad {\rm for} \quad m_a = 5 \,{\rm GeV} \ .
    \label{eq:params2}
\eeqa
\end{itemize}
Thus a generic feature for the explanation of axion explanation for $g-2$ is that we need large axion couplings to muons as well as large axion couplings to electrons or to photons.

\section{Large axion-fermion couplings}
\label{sec:models}

As shown in the previous section, we need very large axion couplings to fermions or photons to explain the muon $g-2$ discrepancy. In this section, we will argue that the models that generate a large coupling between the axion and SM leptons always contain light charged particles with masses around or below 100 GeV, and additional scalars that can both affect the value of $g-2$ itself and potentially mix with the Higgs boson. 

Before discussing axion coupling to fermions, we want to comment on the axion-photon coupling first. A large axion coupling to photons, i.e., $f_a/\left|c_{\gamma\gamma}\right|\lesssim$ (10 - 25) GeV, as is needed for one possible explanation of muon $g-2$ (see Fig.~\ref{fig:massphotoncoupling}), usually requires charged matter with masses about ${\cal O}(10\text{ - }25)$ GeV. Consider the KSVZ type model as an example~\cite{Kim:1979if, Shifman:1979if}. In this type, $a F \tilde{F}$ is generated by integrating out heavy charged vector-like fermions with masses of order $f_a$, assuming that these new fermions have order one charges and PQ charges. The vector-like fermions could be heavier, e.g., 1 TeV and above, if they have charges 7 or larger. Yet requiring the Landau pole of $U(1)_Y$ to be above the Planck scale $\sim 10^{18}$ GeV limits the hypercharge of the heavy matter to be $\lesssim 6$~\cite{Agrawal:2017cmd}. Another difficulty is that these highly charged fermions may not decay quickly since they could only couple to SM particles through very high dimensional operators. This may lead to severe collider and cosmological bounds. For instance, CMS searches have ruled out stable particles with electric charge $\pm 2e$ through Drell-Yan production up to 890 GeV~\cite{CMS:2016ybj}. Thus we will not explore this exotic loophole further. 
One could also consider fermions with large PQ charges (in units of the PQ charge of the scalar which is responsible for PQ breaking). However, the coupling between the PQ scalar and the fermions has to be from high-dimensional operators due to the PQ charges and the masses of the fermions are then exponentially suppressed. More details could be found in Appendix C of Ref.~\cite{Agrawal:2017cmd}.

Another class of possibility to generate a large axion-photon coupling is applying the clockwork mechanism~\cite{Choi:2015fiu, Kaplan:2015fuy, Dvali:2007hz, Choi:2014rja,  Choi:2020rgn} to enhance the field range of the lightest axion beyond its fundamental scale (the PQ breaking scale), in a multiple axion model. It has been used to generate large couplings of photons to a QCD axion~\cite{Farina:2016tgd, Agrawal:2017cmd}.\footnote{The clockwork mechanism has also been used to enhance the couplings of the QCD axion to nucleons and electrons \cite{Darme:2020gyx}.} Yet in all different realizations of the clockwork mechanism, light charged matter is still needed with masses at the fundamental scale, which is about $f_a/c_{\gamma\gamma}$ in our notations; or can lead to dangerous flavor-violating axion couplings.

\subsection{PQ current}
\label{subsec:current}

We briefly review the basic formalism for PQ currents, which we will use later. 
The axion, $a$, is a Nambu-Goldstone boson (NGB) resulting from the spontaneous breaking of a global $\UPQ$. 
Let $\{ \Phi_s \}$, $\{ \psi_f \}$ be a set of scalars and left-handed Weyl fermions with $\UPQ$ charges $\{ q_s \}$ and $\{ q_f \}$, respectively. Then in general the $\UPQ$ current is given by:
\beqa
    J_\mu^{\rm PQ} & = & \sum\limits_s J_\mu^s + \sum\limits_f J_\mu^f \ , \\
{\rm where} \quad    J_\mu^s & = & - q_s \Phi_s^\dagger i \overleftrightarrow{\partial_\mu} \Phi_s \ , \\
    J_\mu^f & = & - q_f \psi_f^\dagger \overline{\sigma}_\mu \psi_f \ .
\eeqa

Substituting the parametrization $\Phi_s = \frac{r_s}{\sqrt{2}} \mathrm{e}^{i \theta_s}$, where $r_s$ and $\theta_s$ are the radial and angular modes of $\Phi_s$ respectively, we note that:
\beq\label{eq:scalar_current}
    -q_s \Phi_s^\dagger i \overleftrightarrow{\partial_\mu} \Phi_s = q_s r_s^2 \partial_\mu \theta_s \ .
\eeq

Goldstone's theorem states that $\langle 0 \vert J^{\rm PQ}_\mu(x) \vert a (p) \rangle = -i f_a p_\mu \mathrm{e}^{-i p\cdot x}$. Thus $J^{\rm PQ}_\mu$ contains $f_a \partial_\mu a$. It can only come from those scalars with VEVs (i.e., those whose radial modes do not annihilate the vacuum: $\langle 0 \vert r_s^2 \vert 0 \rangle = v_s^2 \neq 0$). In models without additional $U(1)$ symmetries under which PQ-charged fields transform, this fixes the linear combination of $\theta_s$ that corresponds to the axion. Defining the canonically normalized $\Phi_s$ phase fields, $a_s \equiv \theta_s v_s$, we obtain:
\beqa
    J_\mu^s & = & f_a \partial_\mu a \ , \label{eq:jpq_scalar} \\
    f_a \, a & \equiv & \sum\limits_s q_s v_s\, a_s \ , \label{eq:axion_def} \\
    f_a^2 & \equiv & \sum\limits_s q_s^2 v_s^2 \ . \label{eq:fa_def}
\eeqa
In models with multiple $U(1)$ symmetries, one must be more careful. For example, in DFSZ-type models~\cite{Dine:1981rt, Zhitnitsky:1980tq}, Higgs fields that carry electroweak charges also carry PQ charge. In this case, there are two Goldstone bosons in the theory, one which is eaten by the $Z$, and one which survives as a light axion. Any linear combination of $U(1)_Y$ and $U(1)_{\rm PQ}$ is a symmetry of the theory, so there is not a unique candidate choice for $J_\mu^\mathrm{PQ}$. Instead, one must take care to identify the light axion as the linear combination of PQ and hypercharge Goldstone bosons which is orthogonal to the combination eaten by the $Z$~\cite{Srednicki:1985xd}.

We are interested in the general properties of models that could generate sizable axion-lepton interactions. The solution to the muon $g-2$ anomaly, as discussed in Sec.~\ref{sec:explanation}, requires sizable couplings between the SM leptons $\ell$ and the axion, of order
\beq\label{eq:g-2_sol}
\bl\vert \frac{f_a}{c_{ii}} \br\vert \sim (25-100)~\GeV \ .
\eeq

There are two possible paths a model could take:
\begin{itemize}
 \item $\ell$ is directly charged under $\UPQ$, or
 \item $\ell$ is PQ-neutral, but coupled to other fields charged under $\UPQ$.
\end{itemize}
After a brief review of fermionic field redefinitions, we devote the rest of this section to studying each of these possibilities in turn.

\subsection{Shifting Yukawa couplings into derivative couplings}
\label{subsec:yuk-to-der}

Derivative couplings of leptons to axions, as in Eq.~\eqref{eq:effective2}, are generated in models of the DFSZ type~\cite{Dine:1981rt, Zhitnitsky:1980tq}, where the Standard Model fermion fields and Higgs bosons carry charges under a PQ symmetry. The general procedure in such a model is to perform a chiral rephasing of a fermion field to shift a Yukawa-type coupling to an axion into a derivative coupling and a coupling of the axion to gauge fields. We start by isolating the phase $\theta_\phi$ of the Higgs field $\phi$ that gives mass to a fermion:
\begin{equation}
y \phi \psi {\tilde \psi} \mapsto m \mathrm{e}^{i \theta_\phi} \psi {\tilde \psi},
\end{equation}
where $\theta_\phi$ is a periodic variable with period $2\pi$ by construction. We then carry out the (well-defined) field redefinition $\psi \mapsto \mathrm{e}^{-i \theta_\phi} \psi$. This has the effect of generating a derivative coupling:
\begin{equation}
\psi^\dagger i {\bar \sigma}^\mu \partial_\mu \psi \mapsto \ldots + (\partial_\mu \theta_\phi) (\psi^\dagger {\bar \sigma}^\mu \psi).
\end{equation}
When $\psi$ carries charge $q$ under a $U(1)$ gauge symmetry, this chiral phase rotation also produces a term
\begin{equation}
\frac{\theta_\phi q^2 \alpha}{4\pi} F_{\mu \nu} {\tilde F}^{\mu \nu}
\end{equation}
in the frame where the gauge field is canonically normalized.

{\em After} carrying out this field redefinition, one can remove the linear combinations of the phase fields $\theta$ from the theory that become massive, and rewrite the low-energy effective Lagrangian in terms of the light axion. This achieves, in a mathematically consistent way, the same results that are often written in the literature using irrational phase rotations of the form $\psi \mapsto \mathrm{e}^{-i c \theta} \psi$ where $c \notin \mathbb{Z}$, which are ill-defined.

\subsection{$\ell$ is PQ-charged}
\label{subsec:e-pq}

To obtain derivative couplings of the leptons to axions in the manner described in Subsec.~\ref{subsec:yuk-to-der}, we need the phase of the Higgs boson giving mass to the leptons to have overlap with the light axion state. DFSZ models accomplish this by assuming the existence of Higgs bosons carrying PQ charge~\cite{Dine:1981rt, Zhitnitsky:1980tq}. We will label these Higgs bosons $H_1$ and $H_2$. These theories also contain a Standard Model singlet field $\Phi$ charged under the PQ symmetry. We assume that there is a scalar potential $V(H_1, H_2, \Phi)$ such that all three scalars get VEVs, and we parametrize the phases of fluctuations around these VEVs as follows:
\begin{equation}
H_1 = \begin{pmatrix} \frac{v_1}{\sqrt{2}}\, \mathrm{e}^{i \theta_1} \\ 0\end{pmatrix}, \quad H_2 = \begin{pmatrix} 0 \\ \frac{v_2}{\sqrt{2}}\, \mathrm{e}^{i \theta_2}\end{pmatrix}, \quad \Phi = \frac{v_\Phi}{\sqrt{2}}\, \mathrm{e}^{i \theta_\Phi}.
\end{equation}
We assume that the potential contains a term of the form $H_1 H_2 \Phi^{\dagger 2} + \mathrm{h.c.}$, which ensures that the PQ charge assignments obey $\mathrm{PQ}(H_1) + \mathrm{PQ}(H_2) = 2 \mathrm{PQ}(\Phi)$, and gives a mass to one linear combination of the phases, $\theta_1 + \theta_2 - 2 \theta_\Phi$. Then the usual Standard Model Higgs VEV is given by $v_\mathrm{EW}^2 = v_1^2 + v_2^2$. However, because the Higgs fields carry PQ charge, the axion decay constant $f_a$ will {\em also} get contributions of order $v_\mathrm{EW}^2$ from the Higgs VEVs. For typical invisible axion models aimed at solving the strong CP problem, this is a non-issue, because $v_\Phi$ is {\em many} orders of magnitude above the electroweak scale, so the Higgs VEVs are a tiny perturbation on the PQ-breaking scale. For our purposes, however, we immediately run into a difficulty. Considering the allowed parameter space in \Eq{eq:params1} and \Eq{eq:params2}, we see that we always have least one coupling (to photons or electrons) suppressed by a scale
\begin{equation}
f_a \sim |c|\,25\,\GeV,
\end{equation}
for $c = c_{\gamma\gamma}$ or $c_{ee}$. That is, we expect $f_a \ll v_\mathrm{EW}$. This means that we should somehow sequester PQ breaking from electroweak breaking, because axion couplings of order $1/v_\mathrm{EW}$ are too small to account for the observed value of $g-2$. The value of $f_a$ also suggests that the radial components of the PQ scalars, $\rho_s \equiv r_s - v_s$, have masses of order $\sim \mathcal{O}(20~\GeV)$, which could also be important for their phenomenology. 

One way to achieve the necessary sequestering is to arrange for a hierarchy in the size of the scalar VEVs, $v_1 \gg v_2, v_\Phi$. Then the Nambu-Goldstone mode eaten by the $Z$ boson will be dominantly $\theta_1$. This leaves two other modes, dominantly contained in $\theta_2$ and $\theta_\Phi$, one combination of which will become the axion. If $v_2, v_\Phi \sim f_a$, then this can be consistent with $f_a \ll v_\mathrm{EW}$. Because the top quark has a large coupling to electroweak symmetry breaking, it should couple to $H_1$. We are interested in obtaining significant lepton couplings to the axion, so we wish the leptons to acquire their mass from $H_2$. This is compatible with either a Type II 2HDM, in which $H_1$ gives mass to up-type quarks and $H_2$ gives mass to down-type quarks and leptons, or to a lepton-specific 2HDM, in which $H_1$ gives mass to {\em all} quarks and $H_2$ gives mass to leptons. Because the phenomenology of the model becomes more complicated when the axion interacts significantly with quarks, we will choose the latter route, in which all axion couplings to quarks will be suppressed by the small ratio $f_a/v_\mathrm{EW}$. For clarity, in the remaining discussion we will denote the Higgs that gives mass to quarks by $H_q$ and the Higgs that gives mass to leptons by $H_l$. We summarize the scalar content of the model in the Table~\ref{tab:dfsz}.
\begin{table}[h!]
\centering
\begin{tabular}{ |c|c|c|c| }
 \hline
 Field & $SU(2)_L$ & $U(1)_Y$ & $\UPQ$ \\ 
 \hline
 $H_l$ & $\mathbf{2}$ & $-\frac{1}{2}$ & $1$ \\
 $H_q$ & $\mathbf{2}$ & $\frac{1}{2}$ & $1$ \\
 $\Phi$ & $\mathbf{1}$ & 0 & 1 \\
 \hline
\end{tabular}
\caption{Scalar sector content: $H_q$ is the Higgs doublet that couples to the SM quarks, $H_l$ is the doublet coupling to the SM leptons, and $\Phi$ is the SM-singlet PQ-charged scalar. }
\label{tab:dfsz}
\end{table}

Here we have used the freedom to choose any linear combination of PQ and hypercharge to assign definite values of $+1$ for the PQ charges of $H_l$ and $H_q$. The  important interactions for our purposes are:
\begin{equation}
V_0(|H_l|, |H_q|, |\Phi|, |H_l H_q|) + \left(\lambda_{ql\Phi} H_l H_q \Phi^{\dagger 2} + y_u H_q Q U^c + y_d H_q^\dagger Q D^c + y_e H_l L E^c + \mathrm{h.c.}\right).
\end{equation}
One could take, for example, $U^c, D^c,$ and $E^c$ to have PQ charge $-1, +1,$ and $-1$ respectively.

The $Z$ boson eats the linear combination $v_q^2 \theta_q - v_l^2 \theta_l$, and (as noted above) the $\lambda_{ql \Phi}$ term gives a mass to the linear combination $\theta_l + \theta_q - 2 \theta_\Phi$. The light axion mode must be orthogonal (in the metric defined by the kinetic terms of the $\theta$ fields) to both of these combinations~\cite{Srednicki:1985xd}. If we define canonically normalized fields $\xi_q = v_q \theta_q$, $\xi_l = v_l \theta_l$, and $\xi_\Phi = v_\Phi \theta_\Phi$, then one can calculate that the light axion mode is
\begin{equation}
a = \frac{1}{f_a} \left(v_\Phi \xi_\Phi  + 2 \frac{v_q v_l}{v_\mathrm{EW}^2}(v_q \xi_l + v_l \xi_q)\right), 
\end{equation}
where $f_a^2 = v_\Phi^2 + 4 v_q^2 v_l^2/v_\mathrm{EW}^2$. In particular, assuming that $v_\mathrm{EW} \approx v_q \gg v_l, v_\Phi$, this reduces to
\begin{equation}
a \approx \frac{1}{\sqrt{v_\Phi^2 + 4 v_l^2}} \left(v_\Phi \xi_\Phi + 2 v_l \xi_l\right) + {\cal O}(v_{l,\Phi}/v_\mathrm{EW}).
\label{eq:lightaxion}
\end{equation}
As promised, the light axion is independent of the mode coupling to quarks up to corrections suppressed by the ratio of small VEVs to the large VEV. Of course, at the level that we are working so far, this axion is {\em massless}, so we must also add some explicit PQ-violating terms to $V$. If these terms involve only $\Phi$ and are relatively small, then we expect that their effects can have little effect on the axion's couplings to Standard Model fields.

We rewrite the fermion couplings in our preferred form by following the recipe outlined in Subsec.~\ref{subsec:yuk-to-der}. In this way, we obtain the following derivative couplings of the light axion:
\begin{align}
{\cal L}_\mathrm{eff} &\supset (\partial_\mu \theta_l) E^{c\dagger} {\bar \sigma}^\mu E^c + (\partial_\mu \theta_q) U^{c\dagger} {\bar \sigma}^\mu U^c - (\partial_\mu \theta_q) D^{c\dagger} {\bar \sigma}^\mu D^c  \nonumber \\
&\mapsto \frac{2}{f_a} \frac{v_q^2}{v_\mathrm{EW}^2} (\partial_\mu a)E^{c\dagger} {\bar \sigma}^\mu E^c + \frac{2}{f_a} \frac{v_l^2}{v_\mathrm{EW}^2} (\partial_\mu a)\left(U^{c\dagger} {\bar \sigma}^\mu U^c - D^{c\dagger} {\bar \sigma}^\mu D^c\right).
  \label{eq:dfsz-deriv-couplings}
\end{align}
In the second line, the $\mapsto$ symbol indicates that we have projected onto only the coupling of the light axion, dropping the contributions that are related to the other, decoupled linear combinations of phases. As expected, we obtain a derivative coupling of the axion to leptons, suppressed by the low scale $f_a$ (because $v_q \approx v_\mathrm{EW}$), while the axion coupling to quarks is further suppressed by the square of the ratio of small to large VEVs.

Notice that, in the absence of any additional Higgs bosons, these couplings are generation-independent: all the leptons obtain masses from the Higgs $H_l$, and so all have the same phase $\theta_l$ that must be rotated away. Hence, the minimal version of this model predicts equal couplings to electrons and muons. Models with additional Higgs bosons could accommodate different couplings, at the cost of adding more possible collider-accessible particles to the theory.

In addition, the chiral rotations we have performed generate couplings of $\theta_l$ and $\theta_q$ to the photon. They do {\em not} generate couplings to the gluons, because the quark fields $U^c$ and $D^c$ were rotated by equal and opposite phases. Taking account of the three generations and the color and charge factors, we obtain a coupling
\begin{equation}
3 (\theta_l + \theta_q) \frac{\alpha}{4\pi} F_{\mu \nu}{\tilde F}^{\mu \nu} \mapsto \frac{6}{f_a}\frac{v_q^2 + v_l^2}{v_\mathrm{EW}^2} \frac{\alpha}{4\pi} a F_{\mu \nu}{\tilde F}^{\mu \nu} = \frac{6}{f_a} \frac{\alpha}{4\pi} a F_{\mu \nu}{\tilde F}^{\mu \nu}.
  \label{eq:dfszcgammagamma}
\end{equation}
Again, the $\mapsto$ indicates projecting onto the light axion. Here we recognize that indeed, $f_a$ plays the role of an axion decay constant, and the coefficient is an integer---reflecting that the light axion combination does, in fact, behave as a periodic field in the low-energy effective theory.

The model makes a correlated prediction for the axion's derivative couplings to leptons and its coupling to gauge fields. In particular, because $E^c$ is a left-handed Weyl fermion, when we compare \Eq{eq:dfsz-deriv-couplings} to \Eq{eq:effective2} we find that $c_{ii}$  has the {\em opposite} sign as $c_{\gamma\gamma}$. However, one could always shift the photon coupling, without affecting the derivative coupling to leptons, by adding additional KSVZ-like fermions, as discussed above.

This model is rife with phenomenological difficulties. Although the DFSZ-like model can generate axion couplings to leptons of the desired form, it comes with a large number of additional, correlated predictions that must also be confronted with data. The theory is a 2HDM, with an additional real scalar singlet from the radial mode of $\Phi$. One must arrange for the VEVs $v_l$ and $v_\Phi$ to be small, without predicting a light charged Higgs boson that would have been discovered at LEP. One must also arrange for the Higgses to be close to the {\em alignment limit} in which the couplings are Standard Model-like, which is difficult when the additional Higgs bosons are so  light. Furthermore, the radial mode of $\Phi$ mixes with the Higgs boson, predicting exotic signals like $h \to aa$ decays. 

The phenomenology of 2HDMs in general has been studied extensively (see, e.g.,~\cite{Branco:2011iw} for a review), and numerous experimental searches have been carried out. The lepton-specific 2HDM discussed here has also been studied under the name of the Type IV 2HDM in Ref.~\cite{Barger:1989fj} and the Type III 2HDM in Refs.~\cite{Craig:2012vn,Craig:2013hca} (readers should be careful, as such terms are not used consistently in the literature). Exotic Higgs decays, and in particular the $h \to aa$ decay with $a \to \gamma\gamma$ or $a \to l^+ l^-$ that can arise in this model, have also been studied extensively (see, e.g.,~\cite{Dobrescu:2000jt, Chang:2006bw, Lisanti:2009uy, Curtin:2013fra}). The constraints from the global electroweak fit could be found in~\cite{Haller:2018nnx}.

Aside from these immediate phenomenological problems, even the original motivation of studying this model is undermined: the axion prediction of the muon $g-2$ in this model is not complete, because the additional Higgs bosons {\em also} couple to the muon and alter the model's prediction for the muon $g-2$. The muon $g-2$ in 2HDM models has been studied extensively (see, e.g.,~\cite{Dedes:2001nx, Gunion:2008dg,  Cherchiglia:2016eui, Cherchiglia:2017uwv}), including in the case of lepton-specific 2HDM models~\cite{Broggio:2014mna, Abe:2015oca, Chun:2015hsa, Chun:2016hzs, Wang:2018hnw, Han:2018znu}. These effects are similar to the model we have discussed, but lack the additional PQ scalar $\Phi$. 

\subsection{$\ell$ is PQ-neutral}
\label{subsec:e-no_pq}

If the SM leptons $\ell$ are not charged under the PQ symmetry, they can only develop couplings to the axion by ``inheriting'' the PQ charges from other degrees of freedom that are charged under $\UPQ$. The simplest way to realize this is by integrating out heavy vector-like fermions, which mix with the SM leptons. 

To generate couplings of axions to muons and electrons, in both electroweak doublets and singlets, we introduce four pairs of vector-like fermions charged under $SU(2)_W \times U(1)_Y \times \UPQ$ (one pair for each flavor in one $SU(2)_W$ representation): 
\beq
\psi_i: (\mathbf{2}, -\frac{1}{2}, 1), \; \; \tilde{\psi}_i: (\mathbf{2}, \frac{1}{2}, -1); \quad \chi_i: (\mathbf{1}, 1, 1), \; \; \tilde{\chi}_i: (\mathbf{1}, -1, -1); \quad i=1,2 \ .
\eeq
In addition, we have a PQ scalar, $\Phi_s$ and the first two generations of SM leptons with the following charge assignments:
\beq
\Phi_s: (\mathbf{1}, 0, 1), \quad L_i: (\mathbf{2}, -\frac{1}{2}, 0), \;\; E^c_i: (\mathbf{1}, 1, 0), \quad i=1, 2 \ . 
\eeq
The Lagrangian contains the following mass terms involving the new fermions: 
\beq
    \mL \supset \sum_{i=e,\mu} - y_{Ei} \tilde{\chi}_i \Phi_s E^c_i - y_{Li} L_i^T \Phi_s \epsilon \tilde{\psi}_i  - M_{i} \chi_i  \tilde{\chi}_i - \Lambda_{i} \psi_i^T \epsilon \tilde{\psi}_i + \text{h.c.}  \ ,
\eeq
where $\epsilon$ is the anti-symmetric $2\times2$ matrix. When the new fermions obtain their masses dominantly from the vector-like mass terms, i.e., $y_{Ei} \langle \Phi_s \rangle \ll M_i$ and $y_{Li} \langle \Phi_s \rangle \ll \Lambda_i$, the equations of motion are, at the leading order,
\beq
\chi_i = - \frac{y_{Ei} \Phi_s E_i^c}{M_i}, \quad \psi_i = - \frac{y_{Li} \Phi_s L_i}{\Lambda_i} \ .
\eeq
Plugging them into the kinetic terms of $\chi$ and $\psi$, after the PQ breaking, we have 
\beq
- \frac{\partial_\mu a}{2f_a} \left( \left|\frac{y_{Ei} f_a}{M_i}\right|^2{E_i^{c}}^\dagger \bar{\sigma}^\mu E_i^c+\left|\frac{y_{Li} f_a}{\Lambda_i}\right|^2{L_i}^\dagger \bar{\sigma}^\mu L_i\right),
\eeq
where we use the parametrization of $\Phi_s$ after PQ breaking: $\Phi_s = \frac{f_a}{\sqrt{2}} e^{i a/f_a}$. Mapping onto the general axion EFT in Eq.~\eqref{eq:effective2}, we have 
\beq
c_{ii} = \frac{1}{2} \left( \left|\frac{y_{Ei} f_a}{M_i}\right|^2 + \left|\frac{y_{Li} f_a}{\Lambda_i}\right|^2\right). 
\eeq 
Note that the axion-lepton coupling is always positive in this case. 
In addition, the vector-like fermions do not generate $c_{\gamma\gamma}$ since they have opposite PQ charges. 

This class of model suffers from several serious problems. First of all, $c_{ee}$ and $c_{\mu\mu}$ always have the same sign and thus they do not provide an explanation for the muon $g-2$ discrepancy, as explained in Sec.~\ref{subsec:parameterspace}. This could be fixed by adding some extra vector-like fermions, which do not have opposite charges under $\UPQ$, to generate a sizable axion coupling to photons.\footnote{Note that for these vector-like fermions leading to a non-vanishing $aF\tilde{F}$ coupling, they could not have a vector-like mass term such as $M \chi \tilde{\chi}$ since they do not have opposite PQ charges. Thus they obtain their masses entirely from coupling to the PQ scalar.} Then we could rely on a combination of large axion-photon coupling and axion-muon coupling to accommodate the observed muon $g-2$. Yet to get large axion couplings, all the vector-like fermions have to be light. More concretely, assuming $y_{E2} = y_{L2} \equiv y$ and $M_2 = \Lambda_2 = M$ for simplicity, we have 
\beq
M \lesssim 500 \, {\rm GeV} \left(\frac{y}{\sqrt{4\pi}}\right)\left(\frac{1/(100 \, {\rm GeV})}{c_{\mu\mu}/f_a}\right) \ ,
\eeq
where we use that $y f_a/\sqrt{2} < M$, the condition for us to integrate out the heavy fermions, and choose the Yukawa coupling close to the perturbative unitarity bound $\sqrt{4\pi}$. These light charged fermions decay quickly to $a$ plus SM leptons (or $W/Z$ plus leptons if those channels are kinematically allowed). Axion subsequently decay to $\gamma\gamma, e^+e^-, \mu^+\mu^-$, resulting in lepton-rich/photon-rich final states. These are highly constrained by the LHC searches. For instance, the vector-like heavy leptons decaying to fully leptonic final states are already ruled out up to 400 GeV using LHC 8 TeV data~\cite{Freitas:2014pua}. More bounds on different decay channels of vector-like leptons could be found in~\cite{Dermisek:2014qca}.

Another issue, which also appears in the DFSZ type model in the previous section, is that these vector-like fermions also lead to a new contribution to the muon $g-2$ in addition to the contribution of the axion loop. For instance, setting only $y_{E2}$ to be non-zero and the other Yukawa couplings to be zero, we have a contribution to  muon $g-2$ from the loop involving $\tilde{\chi}_2$, which is~\cite{Freitas:2014pua} 
\beq
\Delta a_\mu^{\tilde{\chi}} \approx 11 \times 10^{-10} \, y_{E2}^2 \, \left( \frac{100 \, {\rm GeV}}{M_2}\right)^2, 
\eeq
which is comparable to the contribution of the axion loop when the fermion is light. Very recent studies of vector-like leptons plus extended Higgs sector for muon $g-2$ could be found in~\cite{Dermisek:2020cod, Dermisek:2021ajd}.

\section{Conclusions}
\label{sec:con}

A heavy axion-like particle with couplings to leptons and photons provides a tantalizing potential solution to the muon $g-2$ anomaly. In this article, we implemented a full two-loop computation of the Barr-Zee-type diagram, which is valid for all possible mass orderings of the particles in the diagram, and updated the parameter space in the axion EFT framework which could potentially explain the intriguing muon $g-2$ result. As already noted in studies with the previous BNL measurement, the axion couplings have to be large to accommodate the discrepancy between the observed value and the Standard Model prediction for $g-2$. We further investigate simple UV completions to generate such large axion couplings, in particular, large axion-lepton couplings. One generic feature, which arises in different classes of models, is that new light degrees of freedom with masses of order a few 10's to a few 100's GeV have to be present. They could be charged, or neutral but mixed with the Higgs boson, and are thus strongly constrained. In addition, these new states contribute to the muon $g-2$ as well, invalidating the use of the axion EFT alone to study the muon $g-2$ anomaly. 

Our study suggests that to consider an axion's contribution to muon $g-2$, we have to consider more complete models specifying the origins of the axion couplings and other relevant degrees of freedom. Beyond muon $g-2$, there is a relatively less constrained region in the axion mass and coupling plane, for a heavy axion with mass around a few times 10 MeV to 10 GeV. We show that the particle physics models behind this region with large axion couplings are associated with rich phenomenology, which has been probed or could be probed in near-future searches. A more systematic study is beyond the scope of the current paper, but could be worthwhile. 

Lastly, the muon $g-2$ anomaly, the revived long-standing puzzle, could just be among the first (indirect) signals of new physics near the weak scale. More experimental and theoretical efforts are needed to unravel the mystery, e.g., more data from future Fermilab runs and the on-going work at J-PARC~\cite{Abe:2019thb}, as well as more work on understanding the Standard Model hadronic vacuum polarization contributions to $g-2$. Given the possibility of new physics involving the muon, a future high-energy muon collider, which could cover a plethora of new physics signals~\cite{Ali:2021xlw, Han:2020uid, Eichten:2013ckl, Chakrabarty:2014pja, Buttazzo:2018qqp, Bandyopadhyay:2020otm, Han:2021udl, Liu:2021jyc, Costantini:2020stv, Han:2020uak, Capdevilla:2021fmj, Chiesa:2020awd, Han:2020pif} well beyond those related to muon $g-2$~\cite{Capdevilla:2020qel, Buttazzo:2020eyl, Capdevilla:2021rwo, Chen:2021rnl, Yin:2020afe}, deserves serious consideration when planning the future of particle physics.

\section*{Acknowledgments}
We thank Robert Ziegler and Roman Marcarelli for useful comments and corrections.
MBA and JF are supported by the DOE grant DE-SC-0010010 and the NASA grant 80NSSC18K1010.
MR is supported in part by the DOE Grant DE-SC0013607, the Alfred P.~Sloan Foundation Grant No.~G-2019-12504, and the NASA Grant 80NSSC20K0506.
CS is supported by the Foreign Postdoctoral Fellowship Program of the Israel Academy of Sciences and Humanities, partly by the European Research Council (ERC) under the EU Horizon 2020 Programme (ERC-CoG-2015 - Proposal n.~682676 LDMThExp), and partly by Israel Science Foundation (Grant No.~1302/19).

\appendix

\section{Full two-loop results}
\label{app:two-loop}

\begin{figure}[h]
\centering
\includegraphics[width=0.8\textwidth]{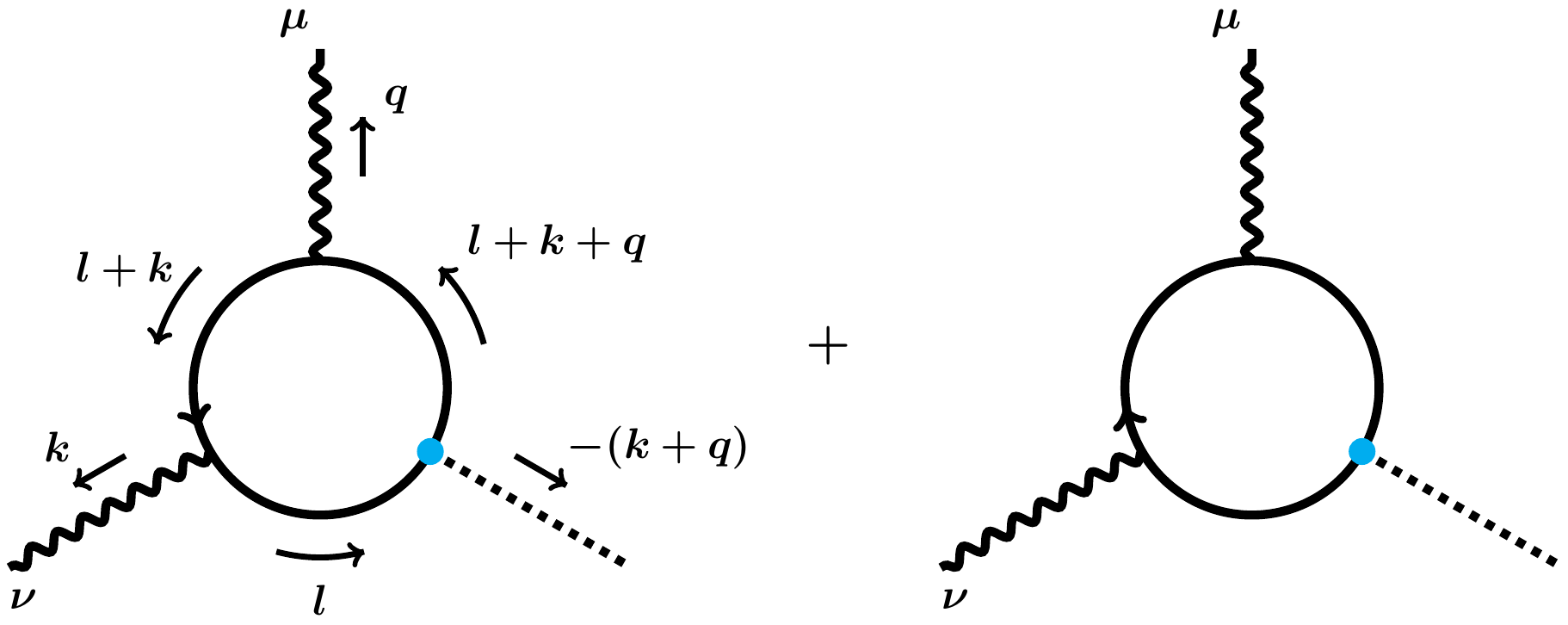}
\caption{Diagrams for the vertex function $\Gamma^{\mu \nu}(k,q,m_i^2)$. }
\label{fig:vertex}
\end{figure}

To calculate the Barr-Zee diagram in the third panel of Fig.~\ref{fig:diagram}, we first compute the contribution of the fermion loop to the three-point vertex function in Fig.~\ref{fig:vertex} with one on-shell photon. We do not require the other photon and the axion to be on mass shell. For the loop calculation, we use Package-X~\cite{Patel:2015tea}. We only keep the term that is linear in $q$, the on-shell photon's momentum. The vertex function is then given by 
\beq
i \Gamma^{\mu \nu} (k,q,m_i^2) = i \frac{c_{ii}\alpha}{\pi f_a} \epsilon^{\mu\nu\alpha\beta}k_\alpha q_\beta  \left( \int_0^1 \mathrm{d}x\,\frac{\Delta^2(x, m_i)}{k^2-\Delta^2(x, m_i)}\right), \; {\rm where} \; \Delta^2(x,m_i) \equiv\frac{m_i^2}{x(1-x)},
   \label{eq:vertexfunction}
\eeq
where $m_i$ is the mass of the fermion running in the loop. The loop integral is ambiguous, shifting by a constant multiple of $\epsilon^{\mu\nu\alpha\beta}k_\alpha q_\beta$ when we shift the loop momentum $l$. This ambiguity in linearly divergent Feynman integrals is familiar from the calculation of the triangle anomaly (see, e.g., Ref.~\cite{Jackiw:1999qq}). The physically correct answer arises from noticing that the derivative coupling in Eq.~\eqref{eq:fermioncoupling} respects a continuous shift symmetry, and hence cannot induce an $a F {\tilde F}$ term in the Lagrangian when massive fermions are integrated out. Indeed, we find that the factor in parentheses in Eq.~\eqref{eq:vertexfunction} approaches a constant, independent of $k^2$, in the limit $m_i \to \infty$. This means that our evaluation of the ambiguous integral has introduced a regularization artifact that breaks the shift symmetry, and we must cancel the symmetry-violating effect with a counterterm. One way to accomplish this is simply to subtract the loop function evaluated in the large mass limit:
\begin{equation}
\Gamma_\mathrm{phys}^{\mu \nu} (k,q,m_i^2) = \Gamma^{\mu \nu} (k,q,m_i^2) - \Gamma^{\mu \nu} (k,q,\Lambda^2),
\end{equation}
where $\Lambda$ is a heavy mass regulator that can be taken to infinity at the end of the calculation.

\begin{figure}[h]
\centering
\includegraphics[width=0.8\textwidth]{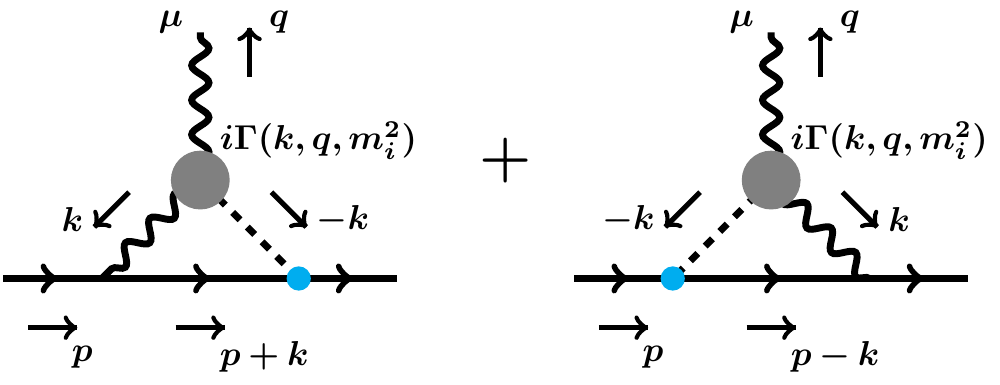} 
\caption{Diagrams for muon $g-2$, inserting the vertex function $i\Gamma^{\mu \nu}(k,q,m_i^2)$. To obtain the correct physical answer, we must subtract equivalent diagrams with an insertion of $i\Gamma^{\mu \nu}(k,q,\Lambda^2)$ in the large $\Lambda$ limit. The internal loop momenta are taken independent of $q$ because we work in the soft $q$ limit, which is sufficient for extracting the magnetic dipole moment}
\label{fig:secondstep}
\end{figure}

Now we insert this vertex function into the diagrams in Fig.~\ref{fig:secondstep}. The final result is 
\beq
\Delta a_\mu^{(3)} =- \frac{c_{\mu\mu}c_{ii} \alpha}{8 \pi^3} \frac{m_\mu^2}{f_a^2} \left[H(m_a, m_\mu, m_i) +  h(m_a,m_\mu, \Lambda) \right],
\eeq
where $H(m_a, m_\mu,m_i)$ is from the diagrams with physical fermions of mass $m_i$ and $h(m_a,m_\mu, \Lambda)$ is the counterterm contribution, from the subtraction of equivalent diagrams with large mass $\Lambda$. The loop function $H$ is given by 
\beqa
H(m_a, m_\mu, m_i) &=& \int_0^1 \mathrm{d}x\, \frac{\Delta^2}{ 2(\Delta^2-m_a^2)}\left[-\frac{2(m_a^2+2m_\mu^2)}{3 m_\mu^2} B(m_\mu^2, m_a, m_\mu)\right. \nonumber \\
&&\left.+\frac{2(\Delta^2+2m_\mu^2)}{3m_\mu^2}B(m_\mu^2,m_\mu, \Delta) + \frac{2(\Delta^2-m_a^2)}{3m_\mu^2}-\frac{m_a^4}{3m_\mu^4} \ln \left(\frac{m_\mu^2}{m_a^2} \right) \right. \nonumber \\
&&\left. + \frac{\Delta^4}{3m_\mu^4} \ln \left(\frac{m_\mu^2}{\Delta^2}\right) \right] 
\label{eq:Hloop}
\eeqa
where $B$ is the function denoted DiscB in Package-X, given by 
\beqa
\label{eq:DiscB}
B(x_0^2, x_1, x_2) &=& \frac{\lambda(x_0^2, x_1^2, x_2^2)}{2x_0^2}\lim_{\epsilon\to 0^+}\int_0^1 \mathrm{d}z\, \frac{1}{x_0^2 z^2 + (-x_0^2-x_1^2+x_2^2)z + x_1^2-i \epsilon}, \nonumber \\
\lambda(a,b,c)&=&a^2+b^2+c^2 -2ab-2ac-2bc. 
\eeqa
The two DiscB function in Eq.~\eqref{eq:Hloop} could be simplified as 
\beqa
B(m_\mu^2, m_a, m_\mu) &=& \frac{m_a \sqrt{m_a^2-4m_\mu^2}}{m_\mu^2} \ln \left(\frac{\sqrt{m_a^2-4m_\mu^2}+m_a}{2m_\mu}\right) \ , \nonumber \\
B(m_\mu^2, m_\mu, \Delta) &=& \frac{\Delta\sqrt{\Delta^2-4m_\mu^2}}{m_\mu^2} \ln \left(\frac{\sqrt{\Delta^2-4m_\mu^2}+\Delta}{2m_\mu}\right) \ . 
\eeqa
The other function $h$ is given by
\beqa
h(m_a, m_\mu, \Lambda)&=& \ln \left(\frac{\Lambda^2}{m_\mu^2}\right) - \frac{m_a^4}{6m_\mu^4}\ln \left(\frac{m_a^2}{m_\mu^2}\right) + \frac{m_a^2}{3m_\mu^2} + \frac{5}{2} \nonumber \\
&+& \frac{m_a \sqrt{m_a^2-4m_\mu^2}(m_a^2+2m_\mu^2)}{6m_\mu^4} \ln \left(\frac{\left(\sqrt{m_a^2-4m_\mu^2}+m_a\right)^2}{4m_\mu^2}\right),
\eeqa
where we ignore higher-order terms of order ${\cal O}(1/\Lambda^2)$ or higher. This is approximately $h \approx \ln \left(\frac{\Lambda^2}{m_a^2}\right) + 2$ in the limit $m_a \gg m_\mu$.

\section{One-loop results}
\label{app:one-loop}

In this appendix, we collect the one-loop results for diagram (1) and (2) in Fig.~\ref{fig:diagram}. These have been computed in Refs~\cite{Bauer:2017ris, Cornella:2019uxs}. We have checked the computations and our results are given by 
\beqa
\Delta a_\mu^{(1)} &=&  -\frac{m_\mu^2 c_{\mu\mu}^2}{16\pi^2 f_a^2} h_1(x), \quad x \equiv \frac{m_a^2}{m_\mu^2} \ , \nonumber \\
\Delta a_\mu^{(2)} &=& -  \frac{m_\mu^2 c_{\mu\mu} c_{\gamma\gamma} \alpha}{8\pi^3 f_a^2} \left(\ln \left(\frac{\Lambda^2}{m_\mu^2}\right) - h_2(x) \right) \ ,
\eeqa
where the loop functions are given by 
\beqa
h_1(x) &=& 1+2x + x(1-x) \ln x + 2x(x-3) \frac{\sqrt{x(x-4)}}{x-4} \ln \left(\frac{\sqrt{x}+\sqrt{x-4}}{2}\right) \ , \nonumber \\
h_2(x) &=& -\frac{5}{2}+\frac{x^2}{6}\ln x - \frac{x}{3}-\frac{x+2}{3} \sqrt{x(x-4)} \ln \left(\frac{\sqrt{x}+\sqrt{x-4}}{2}\right) \ .
\eeqa
Note that our $h_2(x)$ differs from that in Refs~\cite{Bauer:2017ris, Cornella:2019uxs} by a constant. This is because we use the same regularization scheme as in the two-loop calculation in App.~\ref{app:two-loop}, which is different from the regularization scheme in the references. In other words, the different constant could be absorbed by a redefinition of the cutoff $\Lambda$. 

\bibliography{ref}
\bibliographystyle{utphys}
\end{document}